\documentclass[aps,prd,twocolumn,showpacs,nofootinbib,amsmath,amssymb,amsfonts,showkeys]{revtex4-1}
\usepackage{graphicx}
\usepackage{bm}
\usepackage{xcolor}

\begin{document}
\title{Redshifted 21-cm emission signal from the halos in Dark Ages}

\author{B. Novosyadlyj$^{1,2}$, V. Shulga$^{1,3}$, Yu. Kulinich$^{2}$, W. Han$^{1}$}

\affiliation{$^{1}$College of Physics and International Center of Future Science of Jilin University, Qianjin Street 2699, Changchun, 130012, People's Republic of China;}
\affiliation{$^{2}$Astronomical Observatory of Ivan Franko National University of Lviv, Kyryla i Methodia str., 8, Lviv, 79005, Ukraine;}
\affiliation{$^{3}$Institute of Radio Astronomy of NASU, 4 Mystetstv str., 61002 Kharkiv, Ukraine 0000-0001-6529-5610}

\date{21 Nov 2019}
\begin{abstract}
The emission in the hyperfine structure 21 cm line of atomic hydrogen arising in the halos with masses $\sim10^6-10^{10}$ M$_\odot$ from the Dark Ages in the models with Warm Dark Matter (WDM) is analysed.
The halos are assumed to be formed from Gaussian density peaks of cosmological density perturbations at $10\lesssim z\lesssim50$. Semi-analytical modelling of the formation of individual spherical halos in multi-component models shows that gas in them has the kinetic temperature in the range of $60-800$ K under adiabatic compression of the collapsing halo, and the temperature of each halo depends on the time of virialization. It is shown that inelastic collisions between neutral hydrogen atoms are the dominant excitation mechanism for hyperfine structure levels, which pulls the spin temperature closer to the kinetic temperature. The brightness temperature of individual halos is in the range of 1-10 K, depending on the mass of the halo and its virialization redshift, and increasing as these two increase. The apparent angular radii of such halos are in the range 0.06-1.2 arcseconds, their surface number density decreasing exponentially from a few per arcmin$^2$ for the lowest mass and redshift to nearly zero for higher values. Assuming a 1 MHz observation bandwidth  the surface number density of the halo at various redshifts is evaluated as well as beam-averaged differential antenna temperatures and fluxes of hydrogen emission from halos of different masses. The beam-averaged signal strongly depends on the cut-off scale in the mass function of dark ages halos that may be caused by free-streaming of WDM particles. The finding is compared with the upper limits on the amplitude of the power spectrum of the hydrogen 21-cm line fluctuations derived from the recent observation data obtained with MWA and LOFAR.  
\end{abstract}
 

\maketitle

\section{Introduction}

The hyperfine 21 cm line of atomic hydrogen is a very important source of information on the physical conditions, dynamics and distribution of baryon matter on the astrophysical and cosmological space-time scales. Theoretical aspects of its excitation and emission/absorption as well as the experimental ones of its observations are well studied. The eloquently titled review paper by Pritchard and Loeb (2012), ``21 cm cosmology in the 21st Century'' \cite{Pritchard2012}, highlights the current state of knowledge, problems, and prospects for observing the most remote parts of the Universe with special telescopes tuned at the 21 cm hydrogen line redshifted to the meter/decameter wavelengths (see also other excellent review papers by \cite{Barkana2001,Fan2006,Furlanetto2006a,Bromm2011,Galli2013}). The long-term efforts of several scientific groups, aimed at registering a signal from the Dark Ages, have finally given the first results as the EDGES team (Experiment to Detect the Global EoR Signature\footnote{https://www.haystack.mit.edu/ast/arrays/Edges/}) has announced the registration of a 21 cm atomic hydrogen absorption line at $z$=15-20 \cite{Bowman2018}. The brightness temperature about 0.5 K announced by the team, however, does not consist, with the existing predictions of the standard $\Lambda$CDM model. Despite the disputes over registration \cite{Hills2018}, it indicates a certain progress in the observations. The data from other ongoing experiments, e.g., the Murchison Widefield Array\footnote{http://www.mwatelescope.org/} (MWA) and the LOw Frequency ARray\footnote{http://www.lofar.org/} (LOFAR) give an upper limit for the brightness temperature power spectrum of the redshifted 21 cm signal. The MWA observations in the frequency range 75-113 MHz (redshift range $z=12-18$) and LOFAR ones in the frequency range 54-68 MHz (redshift range $z=19.8-25.2$) fix an upper limit of $\Delta^2_{21}<(100\,\rm{mK})^2$ at the comoving scales $k\approx0.5$ h Mpc$^{-1}$ \cite{Ewall-Wice2016} and $\Delta^2_{21}<(121\,\rm{mK})^2$ at $k\approx0.038$  h Mpc$^{-1}$ \cite{Gehlot2018} accordingly.
 
In this paper, we analyse the thermal emission in the hyperfine hydrogen line 21 cm of halos of masses $\sim10^6-10^{10}$ M$_\odot$ formed in the Dark Ages at $z=10-50$, as is expected in the cosmological models with either cold or warm dark matter. Presumably, the first sources of light appeared over that period, meaning the end of Dark Ages, that is why Harker et al. \cite{Harker2012} have reasonably named it 'cosmic dawn'. The galaxy scale halos would turn around, collapse and get virialized during that period, with baryonic matter reheated and becoming a source of thermal emission. The first radiation emitted there may be connected with electron transitions in the hyperfine structure of atomic hydrogen, as its $1\,_1S_{1/2}$ level  is pumped through collisions  when the temperature of the collapsing baryon matter becomes higher than such of the cosmic microwave background (CMB). We will follow the evolution of individual spherical halos which have different initial amplitudes, to evaluate the spin and brightness temperatures, to watch when the absorption changes to emission, to see which are the maximum amplitudes achieved and to understand why they have not been observed up to now. We use the results of our recent analysis of the formation of spherical halos of the masses in the range $10^6-10^{10}$ M$_\odot$ at the end of Dark Ages together with the evolution of their ionic and molecular composition \cite{Novosyadlyj2018}. 

Similar tasks were treated by other authors within other methods. Iliev et al. (2002,2003) \cite{Iliev2002,Iliev2003} used semi-analytical methods and shown that the dense warm minihalos ($10^4-10^8$ M$_\odot$, $T\le10^4$ K) are sources of a 21 cm line radiation that can be seen in emission relative to the CMB. Those authors estimated the brightness temperatures of such halos at different redshifts in the range $z=6-20$ and found them increase with the mass of a halo and the redshift of their virialization. Mini-halos form a forest of redshifted 21 cm emission line. The amplitude of brightness temperature of individual halos ($\delta T_b\sim0.1-0.3$ K) and their angular diameters ($\Delta\Theta\sim0.2''$) make detectable the angular fluctuations in this 21 cm background by ongoing LOFAR and upcoming Square Kilometer Array (SKA) experiments. Using the well-known halo model by Cooray \& Sheth (2002) \cite{Cooray2002} Furlanetto \& Oh (2006) \cite{Furlanetto2006} have shown that minihalos generate reasonably large 21 cm fluctuations. They have computed the power spectrum of 21 cm fluctuations from minihalos and have shown that the signal decreases rapidly as feedback increases the Jeans mass. Shapiro et al. (2006) \cite{Shapiro2006} used the high-resolution gas-dynamic and N-body simulations to predict the 21 cm signal at $z>6$ due to collisional decoupling from the CMB before the UV background is strong enough to make decoupling due to Ly-${\alpha}$ pumping important. They have shown that the 21 cm signal from gas in the virialized minihalos dominates over that from the diffuse shocked gas in the intergalactic medium. 
In \cite{Kuhlen2006} it was used the numerical hydrodynamical + N-body simulations (ENZO) of early structure formation in the $\Lambda$CDM model to investigate the spin and brightness temperatures of the hyperfine line of atomic hydrogen at $z<20$ but before the epoch of cosmic re-reionization. They assumed that gas is heated by X-ray emission from an early miniquasars and shown that collisions between hydrogen atoms can efficiently decouple the spin temperature from the CMB in dense minihalos and the 21 cm signal is strongly enhanced. 

The goal of the paper is the  analysis of the dynamical, thermal, reionization and chemical evolution of spherical halos from initial cosmological perturbations up to virialization into individual halos in Dark Ages, and evolution of their emission in 21 cm line of atomic hydrogen. It will give us the possibility to obtain the dependence of brightness temperature of emission/absorption in this line on the initial amplitude of density perturbation which is a precursor of halo, its mass, the redshift of virialization, kinetic temperature of the gas, etc. We apply the idealisation that is often used in such cases: the initial perturbations are spherically symmetric with a top-hat density profile. In this paper, we suppose initially that kinetic temperature of the gas in halos is adiabatic but later we generalise results for all range of temperatures for which the rate coefficients of collisional excitation/de-excitation are known. Such analysis gives us an understanding which is 21 cm atomic hydrogen emission of halos at the adiabatic stage of their formation before the processes of violent relaxation became important. It can be used in the scenarios which take into account the emissivity of diffuse interstellar gas which is not warmed/cooled yet by non-adiabatic processes. 

The outline of this paper is as follows. In section 2 we describe the models of halos, their physical characteristic, chemical composition, etc. In sections 3 we compute the rate of collisional excitation, the population of hyperfine levels of atomic hydrogen, excitation and brightness temperatures in 21 cm line for halos of different mass and amplitude of initial perturbations. In section 4 we estimate an antenna temperature caused by halos in the field of view of a radio telescope. The discussions and comparisons of the results with observational upper limits on the brightness temperature power spectrum of the redshifted 21 cm signal are presented in section 5. Conclusions are in section 6.
\begin{table*}
\begin{center}
\caption{Physical parameters of halos virialized at different $z_v$: $M_h$ is the total mass of halo, $k$ is the wave number of initial perturbation (seed of halo), $C_k$ is the amplitude of initial curvature perturbation,  $z_v$ is the redshift of virialization, $T_K$ is kinetic temperature of baryonic gas, $\rho_{m}$ is the matter density in virialized halo, $n_{HI}$ is the number density of neutral hydrogen atoms, $n_p,\,n_e$ are the number densities of protons and electrons, $r_h$ is the radius of halo in comoving coordinates, $\theta_h$ is the angular radius of halo.}
\begin{tabular} {ccccccccccc}
\hline
\hline
   \noalign{\smallskip}
$M_h$&$k$&$C_k$&$z_v$&$T_K$&$\rho_{m}$&$n_{HI}$&$n_p\approx n_e$&$r_h$&$\theta_h$\\
 \noalign{\smallskip} 
[M$_{\odot}$]&[Mpc$^{-1}$]&  & &[K] &[g/cm$^3$]&[cm$^{-3}$]&[10$^{-6}$cm$^{-3}$]&[kpc]&['']\\
 \noalign{\smallskip} 
\hline
   \noalign{\smallskip} 
$5.1\cdot10^9$&5&$3.0\cdot10^{-4}$& 29.5&394.5&1.35$\cdot10^{-23}$&0.96 &97.0/3.7&1.83&1.01  \\     \noalign{\smallskip}
              & &$2.5\cdot10^{-4}$& 24.4&293.4&7.79$\cdot10^{-24}$&0.56 &60.3/3.8&2.20&1.03  \\ 
    \noalign{\smallskip}
              & &$2.0\cdot10^{-4}$& 19.3&200.2&3.98$\cdot10^{-24}$&0.28 &33.5/4.0&2.75&1.07  \\  
    \noalign{\smallskip}
              & &$1.5\cdot10^{-4}$& 14.2&122.3&1.67$\cdot10^{-24}$&0.12 &15.4/4.3&3.68&1.12  \\  
    \noalign{\smallskip}
              & &$1.0\cdot10^{-4}$&  9.1& 58.5&4.92$\cdot10^{-25}$&0.035 & 5.1/4.9&5.53&1.24   \\              
 \noalign{\smallskip}
 \noalign{\smallskip}
$6.4\cdot10^8$&10&$3.0\cdot10^{-4}$& 34.5&498.3&2.12$\cdot10^{-23}$&01.51 &142.1/3.6&0.79&0.49\\ 
 \noalign{\smallskip}
              &  &$2.5\cdot10^{-4}$& 28.5&375.0&1.22$\cdot10^{-23}$&0.87 & 88.6/3.7&0.95&0.51\\ 
 \noalign{\smallskip} 
              &  &$2.0\cdot10^{-4}$& 22.6&257.7&6.22$\cdot10^{-24}$&0.44 & 49.4/3.9&1.19&0.52 \\ 
 \noalign{\smallskip}
              &  &$1.5\cdot10^{-4}$& 16.6&158.7&2.60$\cdot10^{-24}$&0.19 & 22.8/4.1&1.59&0.55\\ 
 \noalign{\smallskip}
              &  &$1.0\cdot10^{-4}$& 10.7& 76.4&7.59$\cdot10^{-25}$&0.054 &  7.5/4.7&2.39&0.60\\             
 \noalign{\smallskip}
 \noalign{\smallskip} 
 $8.0\cdot10^7$&20&$3.0\cdot10^{-4}$& 39.3&599.9&3.11$\cdot10^{-23}$&2.22 &195.1/3.6&0.35&0.24\\ 
 \noalign{\smallskip}
               &  &$2.5\cdot10^{-4}$& 32.5&452.7&1.78$\cdot10^{-23}$&1.27 &121.3/3.7&0.42&0.25\\
 \noalign{\smallskip}
               &  &$2.0\cdot10^{-4}$& 25.7&316.8&9.02$\cdot10^{-24}$&0.64 & 67.0/3.8&0.52&0.26\\    
  \noalign{\smallskip}
               &  &$1.5\cdot10^{-4}$& 18.9&192.7&3.72$\cdot10^{-24}$&0.27 & 30.5/4.0&0.70&0.27\\ 
   \noalign{\smallskip}
               &  &$1.0\cdot10^{-4}$& 11.9& 90.4&1.02$\cdot10^{-24}$&0.073 &  9.2/4.4&1.08&0.29\\        
  \hline
\end{tabular}
\label{Tab1}
\end{center}
\end{table*}

All computations in the paper are performed for consistent values of the main parameters of the cosmological model, namely, the Hubble constant $H_0=67.36$ km/s/Mpc, the mean density of baryonic matter in the units of critical one $\Omega_b=0.0494$, the mean density of dark matter $\Omega_m=0.3123$, the mean density of dark energy $\Omega_{de}=0.6877$, its equation of state parameter $w_{de}=-1.028$, spectral index of scalar mode of cosmological perturbations $n_s=0.9666$ and the root mean square of matter fluctuations today in the linear theory $\sigma_8=0.806$ \cite{Planck2018a,Planck2018b}. We used also the value for density contrast at the moment of halo virialization $\Delta_v=178$. The cosmological model with such parameters is practicaly $\Lambda$-model with cold dark matter ($\Lambda$CDM) or warm dark matter ($\Lambda$WDM).

\section{Models of dark ages halos}

The physical conditions in the halos and their chemical composition we obtained by modelling the evolution of individual spherical perturbations in the four-ingredient Universe (cold dark matter, baryon matter, dark energy, and thermal relict radiation) starting from the linear stage at the early epoch, through the quasi-linear stage, turnaround point and collapse. For that we integrate the system of 10 differential equations which describe the evolution of local spherical overdensity and temperature of baryonic gas from the early stage in the radiation-dominated epoch up to collapse stage in dark ages when density of matter reaches the virial value: 
$$\rho_{m}(z_v)=\Delta_v\bar{\rho}_{m}(z_v), \quad \bar{\rho}_{m}(z_v)=\frac{3H^2_0}{8\pi G}\Omega_m(1+z_v)^3.$$ 
The initial conditions are set using the observational data on the power spectrum of the cosmological perturbations. For numerical integration we have designed the computer code which implements the modified Euler method taking into account the derivatives from the forthcoming step and improving the results by iterations \cite{Leveque1998}. Starting from the beginning of the cosmological recombination epoch, to the system of 10 differential equations we attach 3 kinetics equations for recombination of the hydrogen and helium atoms \cite{Seager1999,Seager2000} as well as the system of 19 kinetics equations for formation/distruction of the hydrogen, deuterium and helium molecules \cite{Galli1998,Galli2013}. Since, the complete system of kinetics equations are very stiff, so we use in this part of our code the publicly available 
{\it\large recfast}\footnote{http://www.astro.ubc.ca/people/scott/recfast.html} and {\it\large ddriv1}\footnote{http://www.netlib.org/slatec/src/ddriv1.f} codes. Hence, we evaluate the density of all fundamental components of the Universe (dark matter, dark energy and baryonic matter), the temperature of baryonic gas as well as the abundances of neutral atoms, molecules and their ions in the halos for any time in Dark Ages.
The details of modelling can be found in \cite{Novosyadlyj2016,Novosyadlyj2017,Novosyadlyj2018}. After virialization the internal structure of halos (clumpiness, density profiles of baryonic and dark matter components, kinetic temperature of the gas, their velocity dispersion, reionization and chemical compositions, etc.) crucially depends on the properties of dark components and cosmological parameters. It is studied actively in the last decades using N-body simulations and analytical or semi-analytical methods. The mean profiles of density, temperature, velocity dispersion and abundances of neutral hydrogen and reionization fractions of the baryonic component are different for models with a different type of dark matter and halo formation scenarios (see, for example, \cite{Iliev2001,Wang2009,Tinker2010,Smith2011,Schneider2013,Klypin2016} and citing therein). For estimation of brightness and antenna temperatures of dark ages halos in the redshifted 21 cm line of atomic hydrogen and their dependences on mass and redshift of their virialization we suppose that halos to be homogeneous spheres with top-hat profiles of matter density, kinetic temperature and number density of fractions. 
 
This approach simplifies the problem of a self-consistent solution, and, on the other hand, the obtained results can be easy re-estimated for more realistic models of halo structure.

All physical parameters of halos, which are necessary for computation of the rate of collisional and radiation excitations, the population of hyperfine energy states of atomic hydrogen, excitation and brightness temperatures in 21 cm line, are presented in Table \ref{Tab1} and \ref{Tab1A}.

The mass of halo $M_h$, its radius in comoving coordinates $r_h$ and the wave number $k$ of initial perturbation, which is the seed of halo, are connected by relations
{\small $$\frac{M_h}{M_{\odot}}=152.2\Delta_v(1+z_v)^3\left(\frac{\Omega_mh^2}{0.142}\right)r_h^3
=5.1\cdot10^{9}\left(\frac{\Omega_mh^2}{0.142}\right)\left(\frac{5}{k}\right)^{3},$$}
where $z_v$ is the redshift of virialization of halo, 
$h$ is dimensionless Hubble constant (in units of $100\,\rm{km/s/Mpc}$). Note that angular size of dark ages halos, $\theta_h=r_h/D_A(z)\sim0.1-1$ arcsec, where $D_A(z)$ is the angular diameter distance, is much times smaller the diffraction limit of spatial resolution of up-to-date radio telescopes at frequency of 100 MHz. We present them in the last column of Table \ref{Tab1} and \ref{Tab1A}. They depend on the mass of halos and redshift of its virialization. Since the last dependence is not large we will use in sections 4 and 5 the analytical approximation $\theta_h(M_h)=\alpha M_h^{\beta}$ with best-fit values $\alpha=4.5\cdot10^{-4}$ and $\beta=0.35$.  

After virialization the matter density, kinetic temperature of baryon component and radius of halo do not change. The number density of neutral hydrogen atoms $n_{HI}$ is practically unchanged, while the number densities of protons $n_p$ and electrons $n_e$ are monotonically\footnote{The linear dependence  in the log-log scales is good approximation.} decreased since molecular reactions continue. We present their model values in Table \ref{Tab1} and \ref{Tab1A} at $z_v$ and $z$=10.

\begin{figure}
\includegraphics[width=0.49\textwidth]{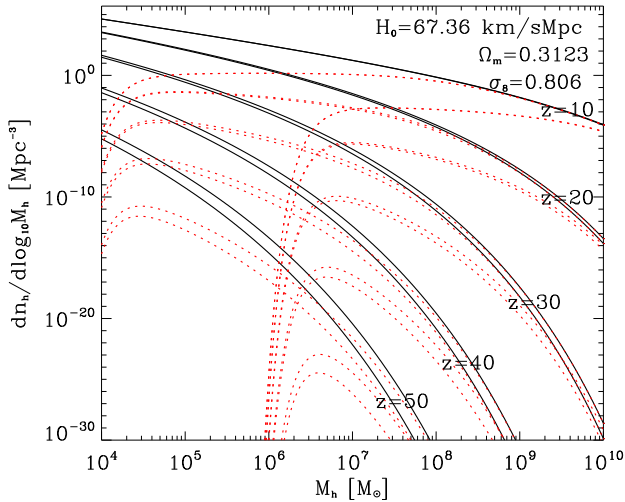}
\caption{The mass functions of halos virialized in Dark Ages for $\Lambda$CDM model (solid lines) and $\Lambda$WDM models (dotted lines with the free-streaming scales at $10^5\,{\rm M}\odot$ (left-hand family of lines) and $10^7\,{\rm M}\odot$ (right-hand family of lines). The double lines correspond to $z \pm \Delta z /2$ with $|\Delta z| $ for 1 MHz band of observations.}
\label{hmf}
\end{figure}

To estimate the number density of halos in Dark Ages we have used the improved Press-Schechter formalism \cite{Press1974,Bond1991} which well matches the mass function of halos at different redshifts obtained in the numerical simulations. The recipes for the semi-analytical computing the mass function of halos in the $\Lambda$CDM and $\Lambda$WDM models can be found in the numerous papers, we follow here the technic proposed in  \cite{Schneider2012,Schneider2013}. We use also the fitting formula for the transfer function of density perturbations power spectrum proposed by \cite{Eisenstein1998}. The $\Lambda$CDM and $\Lambda$WDM linear power spectra of mater density perturbations normalized by computation of $\sigma_8$ with top-hat filter well match the linear power spectrum inferred from different cosmological probes which is presented in Figure 19 of \cite{Planck2018a}. We rescaled its amplitude at the redshifts of Dark Ages by the square of the growth function $D(z)$ proposed in \cite{Carroll1992}. The mass functions of halos computed with Sheth-Tormen's function of the first crossing distribution \cite{Sheth1999} are shown in Figure \ref{hmf}. The double lines correspond to $z\pm \Delta z/2$ with $\Delta z=(1+z)^2\Delta\nu_{obs}/\nu_0$, where $\Delta\nu_{obs}$ is frequency band of observations which we assume here  equal 1 MHz. They are visually indistinguishable from each other at $z$=10. Solid lines show the mass functions at different redshifts in the $\Lambda$CDM model, while dotted lines in the $\Lambda$WDM models. The last ones are obtained by multiplying of $\Lambda$CDM mass functions on the suppression factor 
$(1+M_{hm}/M_h)^{-\alpha}e^{(-M_{fs}/M_h)^2}$ with $\alpha \simeq 1.16$ \cite{Schneider2012}. The free-streaming mass-scale $M_{fs}$ and half-mode mass scale $M_{hm}$ are expressed as functions of the mass of the thermal relic WDM particle  $m_{WDM}$ 
as follows: 
\begin{eqnarray}
M_{fs}&\simeq& 4.8\cdot 10^6 \left(\frac{m_{WDM}}{1\, \textrm{keV}}\right)^{-3.33}~\, h^{-1}M_{\odot} \nonumber\\
M_{hm}&\simeq& 1.3\cdot 10^{10}\left(\frac{m_{WDM}}{1\, \textrm{keV}}\right)^{-3.33}~\, h^{-1}M_{\odot}
\label{mwdm}
\end{eqnarray}
(see also Fig. 1 in \cite{Schneider2012}). The suppression factor describe how small halos are erased in the WDM models when they are below the free-streaming scale, $M_h<M_{fs}$, and delay the growth of larger ones in the range of masses $M_{fs}<M_h<M_{hm}$.  
Thus, the halo statistics resulted by observations gives a constraint on the mass of dark matter particles. In particular, the luminosity functions of faint galaxies obtained from the Hubble Ultra Deep Field for different redshifts are used to constrain the WDM particle mass. So, the constraint $m_{WDM}$>1.3 keV (2$\sigma$) was obtained using data on galaxies at $z$=6-8 \citep{Schultz2014}, $m_{WDM}$>1.8 keV(2$\sigma$) using data on galaxies at $z$=2 \citep{Menci2016}, $m_{WDM}\ge1.5$~keV using data on galaxies at $z$=6, 7 and 8 \citep{Corasaniti2017}, $m_{WDM}$>0.9 keV (2$\sigma$) using data on   two galaxies at $z$=10 \citep{Pacucci2013}, etc.  
Analysis of gravitational wells of the Milky Way Satellites gives the constraints on the WDM particle mass: $m_{WDM} \ge 2$ keV \citep{Lovell2012,Kennedy2014} and $m_{WDM}$ >  2.3keV \citep{Polisensky2011}, and for the galaxies of the Local Group, $m_{WDM} \ge 1.8$~keV \citep{Horiuchi2014}. Constraints on the WDM particle mass from high-redshift long gamma-ray bursts give $m_{WDM}$>1.6-1.8 keV at 95\% CL \citep{Souza2013} while constraints from Ly-$\alpha$ forest data give $m_{WDM}$>5.3 keV (2$\sigma$) \citep{Viel2013,Irsic2017}.  

However, as it is shown in \cite{Sekiguchia2014}, more promising constraints expected from the observations of 21 cm line from halos at $z$>5, which are sensitive to the WDM particle mass up to tens keVs. In this paper, we focus on the halos at $z\gtrsim$10. 
  
\section{Spin and brightness temperatures of halos}
\begin{figure*}
\includegraphics[width=0.49\textwidth]{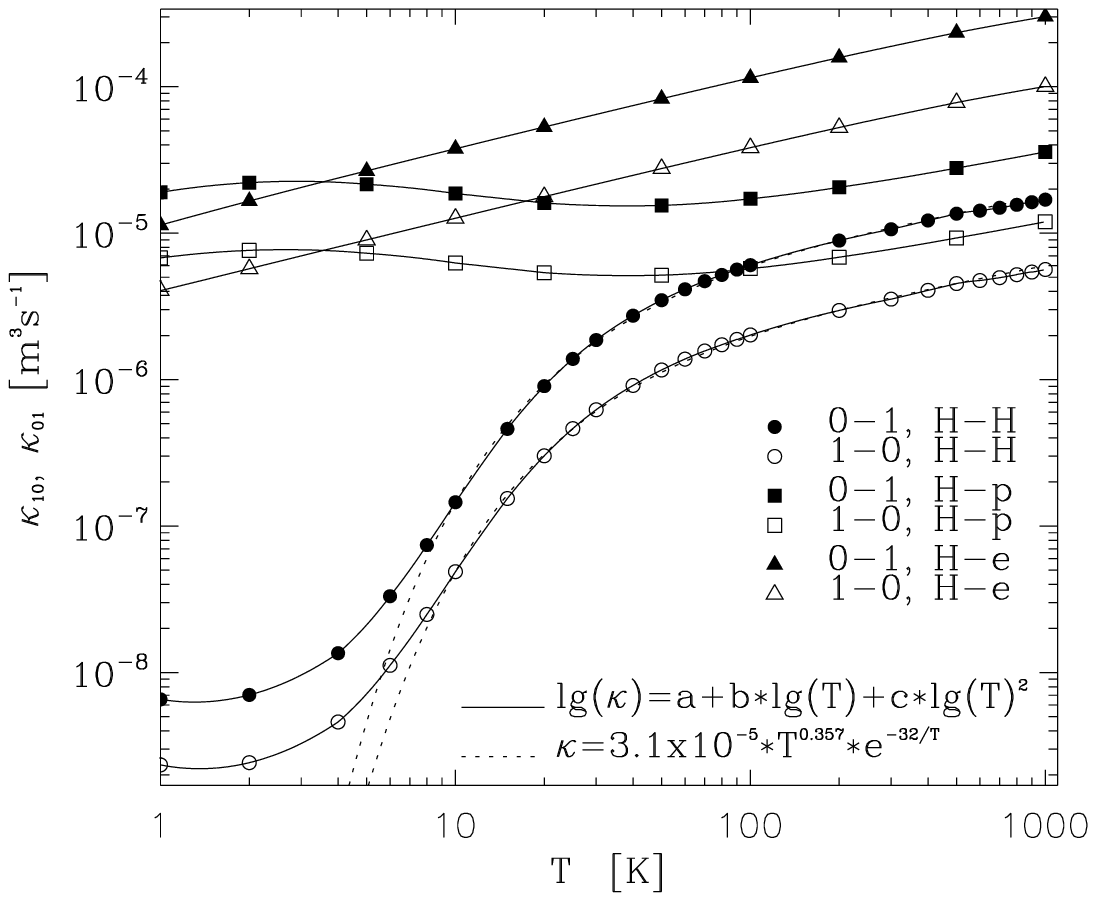}
\includegraphics[width=0.49\textwidth]{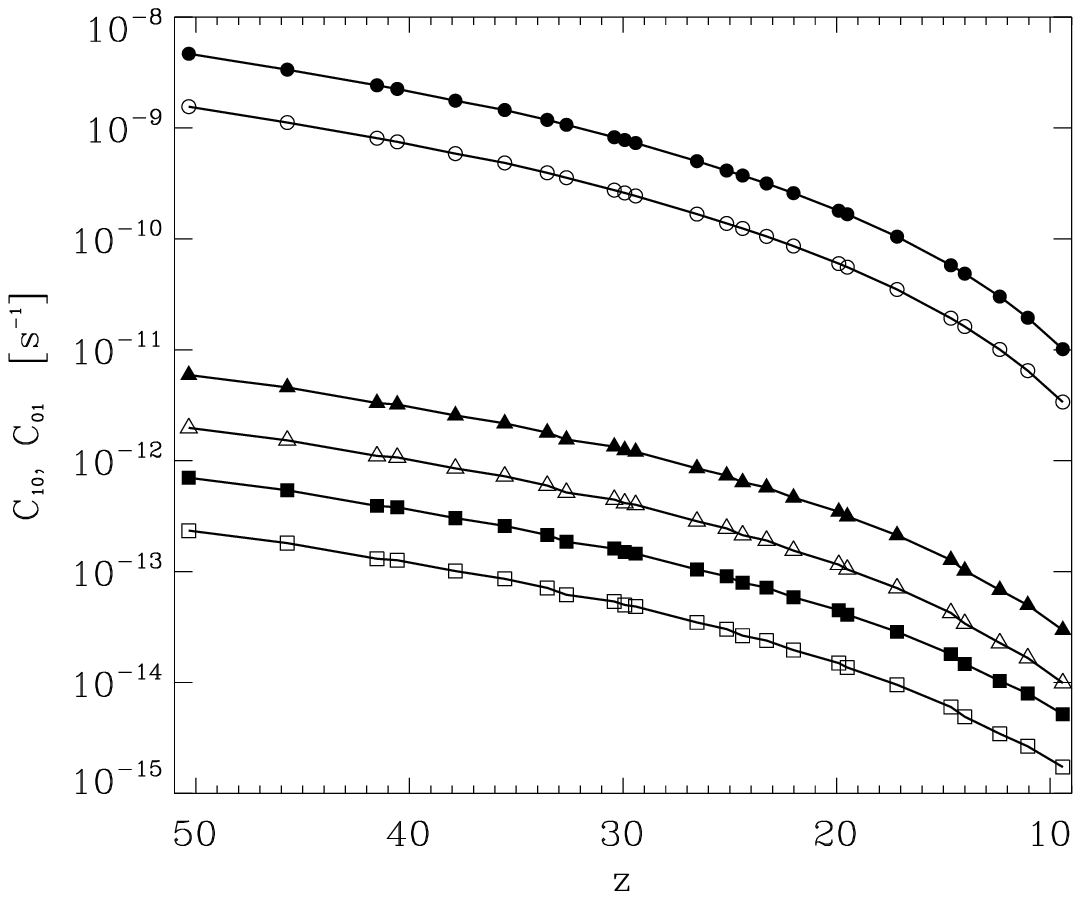}
\caption{Left: The temperature dependences of rate coefficients for collisional excitation ($\kappa_{01}$, filled symbols) and de-excitation ($\kappa_{10}$, open symbols) of the hyperfine line of atomic hydrogen by H \cite{Allison1969,Zygelman2005}, e \cite{Furlanetto2007a} and p \cite{Furlanetto2007b}. The thin solid lines are our approximations, and the dotted lines are approximation by \cite{Kuhlen2006}. Right: The rate of collisional excitation $C_{01}=\kappa_{01}^Xn_X$ (filled symbols) and de-excitation $C_{10}=\kappa_{10}^Xn_X$ (open symbols) in halos virialized at $z=50-10$.}
\label{kappa_H21}
\end{figure*}
\begin{figure}
\includegraphics[width=0.49\textwidth]{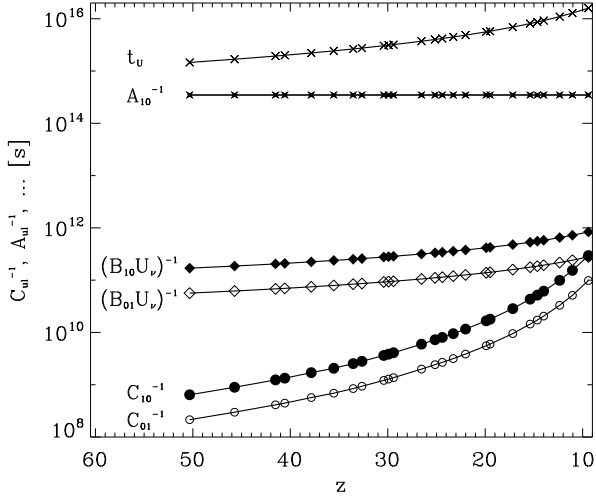}
\caption{The inverse rates of excitation/de-excitation $C_{10}^{-1}$, $C_{10}^{-1}$, $(B_{10}U_{\xi_{10}})^{-1}$, $(B_{01}U_{\xi_{10}})^{-1}$, and $(A_{10})^{-1}$ in units of s in halos virialized at $z$=30.35, 25.08, 19.81, 14.56, and 10.34. The age of the Universe is shown by the line with crosses. }
\label{h21_time}
\end{figure}

Since \, the\, energy\, of\,\, hyperfine\,\, structure\,\,\, transition $1_{0}S_{1/2}\rightarrow1_{1}S_{1/2}$ is $E_{10}=0.068\,\rm{K}\ll T_{CMB}$, the cosmic microwave background is important for pumping of the hyperfine upper level of atomic hydrogen. When CMB pumping is dominant mechanism of excitation and de-excitation of this level the atomic hydrogen is not visible on the CMB background in the redshifted 21 cm line, the spin temperature in this case is equal to the radiation one, $T_s=T_{CMB}$.    
But when other mechanisms, collisions or UV radiation, are competitive the atomic hydrogen can be observed in the absorption or emission of CMB quanta, depending on the relation of the spin temperature to radiation one. The baryonic matter in the halos formed long before the re-reionization are warmed adiabatically at the stage of contraction to a temperature $T_K>T_{CMB}$ and by shock waves at the late stages of virialization to a temperature $T_K\gg T_{CMB}$. The collision pumping decouples the spin temperature from radiation one, $T_s>T_{CMB}$, and atomic hydrogen emit the extra quanta of hyperfine transition. Later, when X-ray and UV radiation of  the first sources (shock waves,  stars and quasars) will appear the additional heating and excitation/de-excitation mechanisms will initially enhance and then suppress the luminous of halos in the redshifted 21 cm line \cite{Shapiro2006,Kuhlen2006,Pritchard2012}. In this paper we restrict ourselves to the analysis of the regions in which the Ly-$\alpha$ pumping is negligible even after sources turn on at $z<20$. In other words, we  only consider collisional pumping in the halos before the luminous sources appear. This approximation simplifies the computation of the spin temperature essentially:  the two-level model of atom is a good approximation since energies of upper level ($n\ge2$) excitation are essentially larger than energy of perturbers in dark ages halos.     
The main equation which describes the evolution of populations of levels $1_{0}S_{1/2}$ and $1_{1}S_{1/2}$ is as follows \cite{Loeb2004}:
\begin{eqnarray}
\label{dy2dz}
&&\hskip-0.5cm\frac{dy}{dz}=\left[y(B_{01}U_{\nu}+C_{01})\right. \\
&&\left.-(1-y)\left(A_{10}+B_{10}U_{\nu}+C_{10}\right)\right]/H(z)(1+z),\nonumber
\end{eqnarray}
where $y\equiv n_0/n_H$ is fraction of atomic hydrogen on the bottom level and $U_{\nu}=8\pi h_P\nu^3c^{-3}\left[\exp{(h_P\nu/k_BT_{CMB})}\right]^{-1}$ is the energy density of CMB radiation at the frequency of hyperfine transition $\nu$. The Einstein coefficient of spontaneous transition, radiation induced one and absorption of radiation are connected by known relations: $A_{10}=8\pi h_P\nu^3c^{-3}B_{10}=2.875\cdot10^{-15}$ s$^{-1}$, $B_{01}=3B_{10}$. We present the total collision rate of excitation $C_{01}$ and de-excitation $C_{10}$ as the sum of excitation/de-excitation by different perturbers, i.e. neutral hydrogen, protons and electrons, that are most abundant in the dark ages halos: 
\begin{eqnarray}
C_{01}&=&\kappa_{01}^Hn_H+\kappa_{01}^pn_p+\kappa_{01}^en_e,\nonumber \\
C_{10}&=&\kappa_{10}^Hn_H+\kappa_{10}^pn_p+\kappa_{10}^en_e\nonumber.
\end{eqnarray}
Here $n_H$, $n_p$ and $n_e$ are the number densities of neutral hydrogen, protons and electrons in the halos. The coefficients of collision de-excitation $\kappa_{10}^H$, $\kappa_{10}^p$ and $\kappa_{10}^e$ we took from the papers \cite{Allison1969,Zygelman2005,Furlanetto2007a,Furlanetto2007b} accordingly, where they are tabulated as a function of temperature. The coefficients of collisional excitation we computed as $\kappa_{01}^i=3\kappa_{10}^i\exp{(-h_P\nu/k_BT_K)}$. They are presented in the left panel of Figure \ref{kappa_H21}. 
In the right panel we show the rates of collisional excitation $C_{01}^i=\kappa_{01}^in_i$ and  de-excitation $C_{10}^i=\kappa_{10}^in_i$ by each perturber in the conditions ($n_i$ and $T_K$) of halos which we will analyse below. One can see that main perturber in the dark ages halos is neutral hydrogen since its number density is $\sim10^4$ times larger. It is interesting to compare them with rates of spontaneous and induced de-excitations as well as with excitation by CMB radiation. In order to compare all of them with the rate of change of the temperature of CMB radiation caused by expansion of the Universe we present their inverse values in Figure \ref{h21_time}. We see that the fastest process is the collisional excitation of upper hyperfine level, the slowest one is the spontaneous transition to the bottom level. This is the physical base of thermal emission. The fact that all processes are faster than the rate of expansion of the Universe means that the quasi-stationary solution of equation (\ref{dy2dz}) 
\begin{eqnarray}
n_0=n_H\frac{A_{10}+B_{10}U_{\nu}+C_{10}}{A_{10}+B_{10}U_{\nu}+C_{10}+B_{01}U_{\nu}+C_{01}}, \nonumber \\
n_1=n_H\frac{B_{01}U_{\nu}+C_{01}}{A_{10}+B_{10}U_{\nu}+C_{10}+B_{01}U_{\nu}+C_{01}},\nonumber 
\end{eqnarray}
can be a good approximation for numerical one. Substitution it in the Boltzmann-like relations $n_1/n_0=3e^{-h_P\nu/k_BT_s}$ gives us the well-known expression for the spin temperature \cite{Field1959}
\begin{equation}
T_s=T_K\frac{T_{CMB}+T_0}{T_K+T_0}, \quad T_0=\frac{h_P\nu}{k_B}\frac{C_{10}}{A_{10}}.
\label{Tsa}
\end{equation}

\begin{figure}
\includegraphics[width=0.49\textwidth]{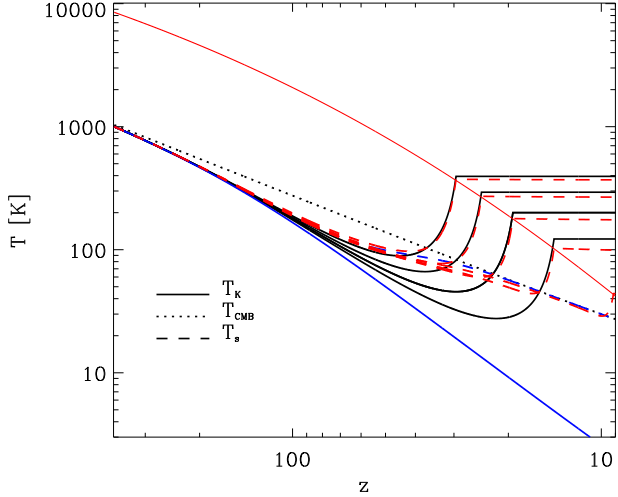}
\caption{Evolution of the adiabatic temperature of baryonic gas  (solid black lines) and spin temperature of neutral hydrogen atoms (dashed red lines)  in the halos with mass $M_h=5.1\cdot10^9$ M$_{\odot}$ which virializes at $z_v$=29.5, 24.4, 19.3, 14.2 and 9.1 (solid lines from top to down). The thin solid red line is the analytical approximation of spin temperature in the halos virialized at $z$. The dotted line shows the temperature of the cosmic microwave background radiation, the lowest solid blue line shows the adiabatic temperature of baryonic gas and the dashed blue line shows the spin temperature of neutral hydrogen atoms at the cosmological background.}
\label{Tex}
\end{figure}

We integrate the equations (\ref{dy2dz}) jointly with kinetic equations of formation/distruction of molecules and evolution of density and velocity perturbations of all ingredients. The results are presented in Figure \ref{Tex} and Table \ref{Tab2}. 
One can see that  at $z\gtrsim100$ the spin temperature follows a kinetic temperature of baryonic matter. Later, at the cosmological background where $T_K<T_{CMB}$ via adiabatic cooling the collision excitations becomes ineffective because of decreasing of the number density of perturbers the spin temperature (blue dashed line in Figure \ref{Tex}) decouples from the kinetic temperature (lowest solid blue line) and approaches to the radiation one (dotted line) at $z\approx25$. In the halos the spin temperature (red dashed lines) follows a kinetic one when $T_K\gtrsim T_{CMB}$. In all cases the quasi-stationary solution (\ref{Tsa}) do not differ from numerical one more than tiny parts of percents.

\begin{table*}
\begin{center}
\caption{Parameters of halos as emitters of 21 cm line.}
\begin{tabular} {cccccccccc}
\hline
\hline
   \noalign{\smallskip}
$M_h$&$C_k$ &$\nu_{obs}$&$\Delta\nu_{fr}$&$T_s$&$\tau_{\nu}^0$&$\langle\delta T_{br}\rangle$&$\left(\frac{d\delta F}{d\nu}\right)_{obs}$&$\theta_{hwhm}$&$\sigma_h $\\
 \noalign{\smallskip} 
[M$_{\odot}$]& &[MHz]&[kHz]&[K]& &[K]&[$\rm{nJy}$]&[arcsec]&[arcmin$^{-2}$]\\ 
 \noalign{\smallskip} 
\hline
   \noalign{\smallskip} 
$5.1\cdot10^9$ &$3.0\cdot10^{-4}$& 46.5&  0.660&  373.5&    3.77&    8.32&   39.89&    0.97&  4.95$\cdot10^{-27}$ \\ 
    \noalign{\smallskip}
               &$2.5\cdot10^{-4}$& 55.9&  0.683&  271.7&    4.16&    7.12&   51.67&    1.00&  2.42$\cdot10^{-19}$ \\ 
    \noalign{\smallskip}
               &$2.0\cdot10^{-4}$& 69.9&  0.706&  179.2&    4.88&    5.60&   68.05&    1.04&  5.58$\cdot10^{-13}$ \\  
    \noalign{\smallskip}
               &$1.5\cdot10^{-4}$& 93.4&  0.737&  102.7&    6.12&    3.81&   91.48&    1.09&  6.49$\cdot10^{-8}$ \\  
    \noalign{\smallskip}
               &$1.0\cdot10^{-4}$&140.3&  0.766&   44.0&    9.13&    1.58&  103.27&    1.21&  3.54$\cdot10^{-4}$ \\         
 \noalign{\smallskip}     
 \noalign{\smallskip} 
$6.4\cdot10^8$ &$3.0\cdot10^{-4}$& 40.0&  0.638&  478.4&    1.77&    7.13&    6.11&    0.46&  4.95$\cdot10^{-24}$\\ 
 \noalign{\smallskip}
               &$2.5\cdot10^{-4}$& 48.1&  0.665&  353.6&    1.91&    6.36&    8.23&    0.47&  5.82$\cdot10^{-17}$\\ 
 \noalign{\smallskip} 
               &$2.0\cdot10^{-4}$& 60.2&  0.690&  236.3&    2.20&    5.35&   11.52&    0.49&  4.00$\cdot10^{-11}$\\ 
 \noalign{\smallskip}
               &$1.5\cdot10^{-4}$& 80.5&  0.724&  137.9&    2.69&    4.04&   17.06&    0.52&  1.76$\cdot10^{-6}$\\ 
 \noalign{\smallskip}
               &$1.0\cdot10^{-4}$&121.4&  0.758&   59.7&    3.93&    2.10&   23.86&    0.57&  4.68$\cdot10^{-3}$\\               
 \noalign{\smallskip}
 \noalign{\smallskip}
$8.0\cdot10^7$ &$3.0\cdot10^{-4}$& 35.2&  0.616&  580.7&    0.86&    4.94&    0.80&    0.22&  3.20$\cdot10^{-21}$\\ 
 \noalign{\smallskip}
               &$2.5\cdot10^{-4}$& 42.4&  0.643&  432.6&    0.91&    4.51&    1.10&    0.23&  1.06$\cdot10^{-14}$\\
 \noalign{\smallskip}
               &$2.0\cdot10^{-4}$& 53.2&  0.676&  295.3&    1.02&    3.97&    1.61&    0.23&  2.63$\cdot10^{-9}$\\
 \noalign{\smallskip}
               &$1.5\cdot10^{-4}$& 71.5&  0.708&  171.8&    1.24&    3.21&    2.57&    0.25&  4.88$\cdot10^{-5}$\\
 \noalign{\smallskip}
               &$1.0\cdot10^{-4}$&110.0&  0.747&   72.8&    1.80&    1.95&    4.35&    0.27&  7.53$\cdot10^{-2}$\\   
  \hline  
\end{tabular}
\label{Tab2}
\end{center}
\end{table*}

The spin temperatures for virialized halos are close to their kinetic temperatures (in our case adiabatic) which are defined by the redshift of virialization. So, we can approximate it by simple expression
\begin{equation}
T_s=a\cdot10^{-\left[b/(1+z)\right]^c},
\label{Ts_appr}
\end{equation}
where the best-fit values of coefficients $(a,\,b,\,c)$ are as follows $(3.1769\cdot10^5,\,2.3468\cdot10^3,\,0.24832)$. It is shown in the Figure \ref{Tex} by thin solid red line. 

Figure \ref{Tex} helps us to understand how and when the emission of virialized halos and absorption of inter halo medium (IHM) in the redshifted 21 cm line of the hydrogen atoms arise in the case when they are excited by CMB photons and collisions only. In the IHM the minimal value of $\mathcal{T}\equiv(T_s-T_{CMB})/T_{s}$ about -0.55 is reached at $z\sim74$ (blue dashed and dotted lines in the figure). In the halo virialized at the same redshift this ratio is $\mathcal{T}_h\approx0.86$ (thin solid red and dotted black lines). The same values for $z$=20  are $\mathcal{T}_{IHM}\approx-0.009$ and $\mathcal{T}_h\approx0.7$, for $z=10$  they are $\mathcal{T}_{IHM}\approx-0.0007$ and $\mathcal{T}_h\approx0.4$, accordingly. It can mean that the effect of emission by separate virialized halos is main in the interested range of redshift.

Now we can estimate the intensity of emission in the hyperfine line of atomic hydrogen from dark ages halos. For that the standard radiative transfer equation 
$$dI_{\nu}/ds=-\alpha_{\nu}I_{\nu}+\epsilon_{\nu}$$ 
must be solved, where $I_{\nu}$ is specific intensity per unit frequency $\nu$, $s$ is the pass along the sight of view in the halo, $\alpha_{\nu}$ and $\epsilon_{\nu}$ are absorption and emission coefficients. For homogeneous isothermal halo at the CMB background it has the simple analytic solution
\begin{equation}
I_{\nu}=\left(\frac{\epsilon_{\nu}}{\alpha_{\nu}}\right)_h(1-e^{-\tau_{\nu}})+I_{\nu}^{cmb}e^{-\tau_{\nu}},\nonumber
\end{equation}
where $\tau_{\nu}\equiv\alpha_{\nu}s$ is optical thick of halo or its opacity in a point of observation. Since the hyperfine frequency $\nu$ is in the  
Rayleigh-Jeans range of radiation energy distribution it is comfortable to use a brightness temperature $T_{br}$ instead intensity: $I_{\nu}=2k_BT_{br}\nu^2/c^2$.
Since the useful signal is the difference of intensities $\delta I_{\nu}^h=I_{\nu}^h-I_{\nu}^{cmb}$ in any point of halo we obtain the final expression for differential brightness temperature at the observable frequency $\nu_{obs}$ \cite{Pritchard2012} 
\begin{equation}
\delta T_{br}=\frac{T_s-T_{CMB}}{1+z}(1-e^{-\tau_{\nu}})
\label{e_dTbr}
\end{equation}
Up to now we discussed the hyperfine line emission as mono\-chro\-matic. In the reality in the rest frame of halos the line is extended by thermal motion and collisions of hydrogen atoms with each other and other atoms, ions and molecules. The width of line at a half-maximum caused by the thermal Doppler broadening is as follows \cite{Lang1974}
\begin{equation}
\frac{\Delta\nu(z)}{\nu_0}=\sqrt{\frac{8\ln{2}\,k_B T_K(z)}{c^2m_H}}.
\label{Dnuz}
\end{equation}
It is taken into account in the absorption coefficient 
\begin{equation}
\alpha_{\nu}\equiv n_0B_{01}\phi(\nu)=\frac{3c^2}{8\pi\nu^2}\frac{n_0}{\Delta\nu}\left[1-e^{-h\nu/k_B T_s}\right]A_{10}.
\label{alpha}
\end{equation}
For an observer the opacity is maximal in the center of halo, $\tau_{\nu}^0=2\alpha_{\nu}r_h$ (Table \ref{Tab2}), and decreases to zero at the edge as
\begin{equation}
\tau_{\nu}=\tau_{\nu}^0\sqrt{1-\theta^2/\theta_h^2},
\label{tau}
\end{equation}
where $\theta$ is the angle between directions to the center and to the observable point on the halo, $\theta_h$ is the angular radius of the geometrically limited halo (Table \ref{Tab1} and \ref{Tab1A}). 
\begin{figure*}
\includegraphics[width=0.32\textwidth]{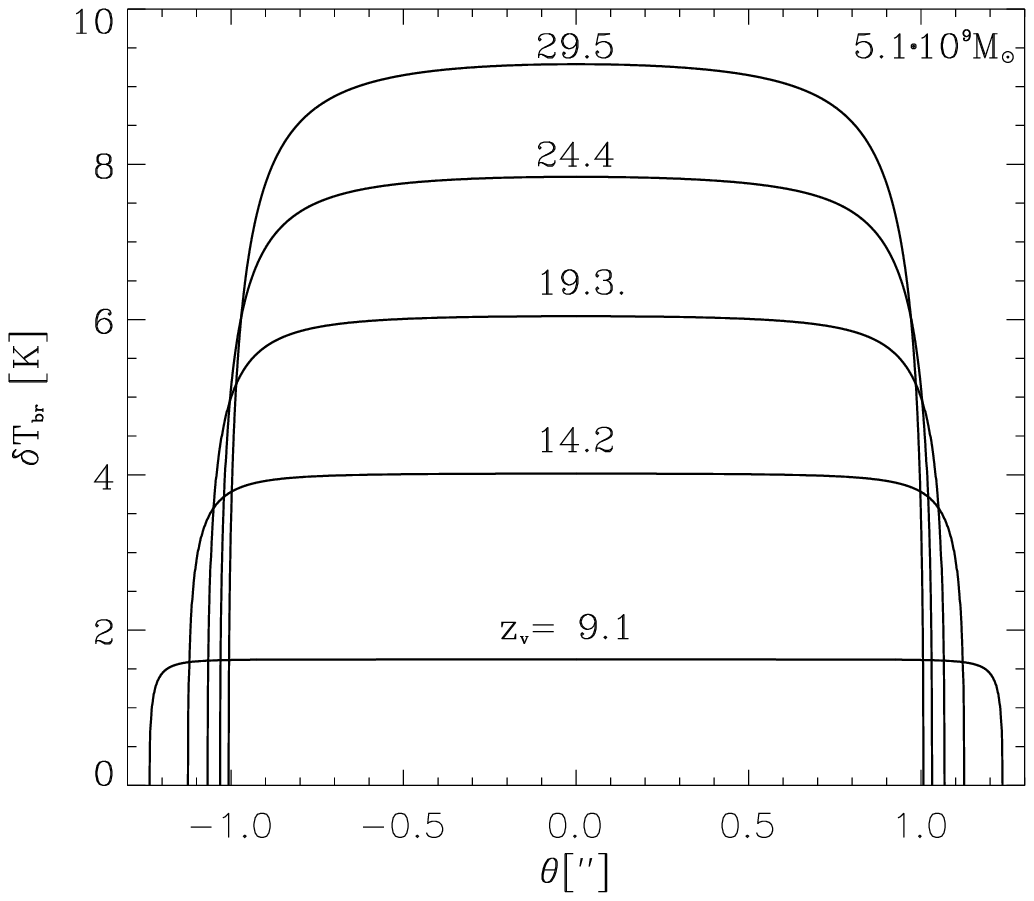}
\includegraphics[width=0.32\textwidth]{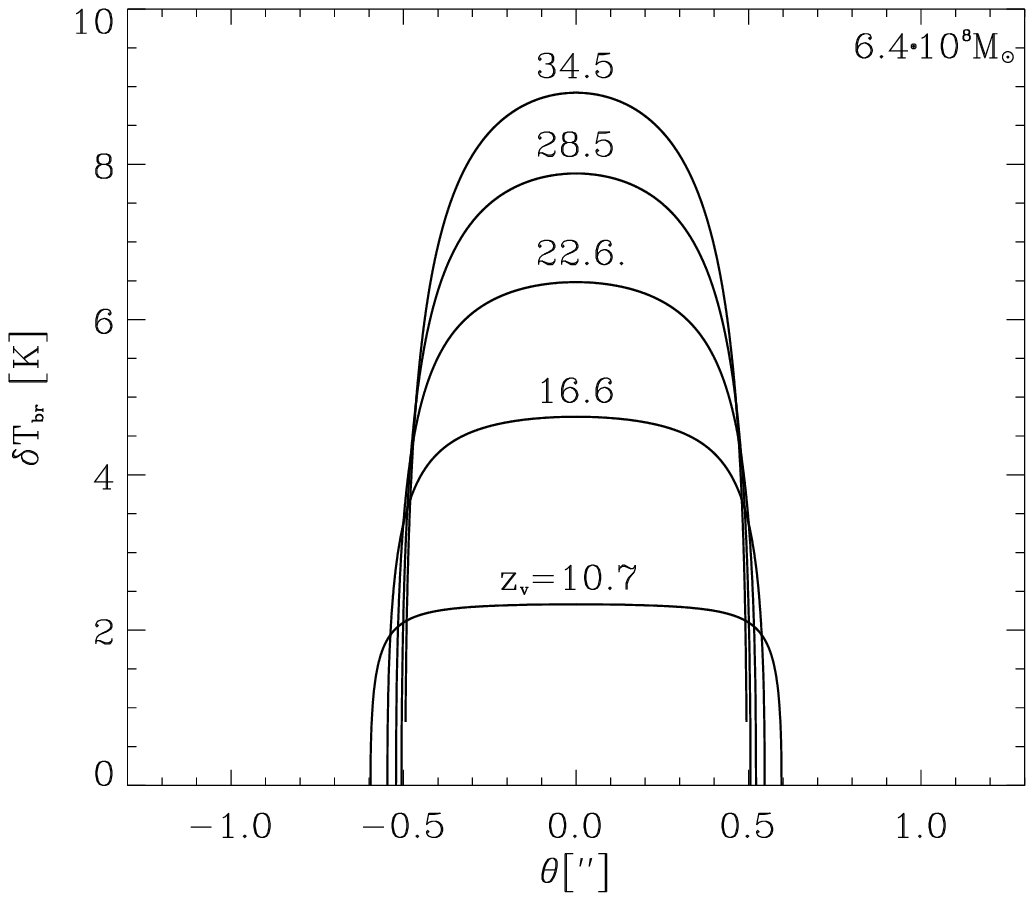}
\includegraphics[width=0.32\textwidth]{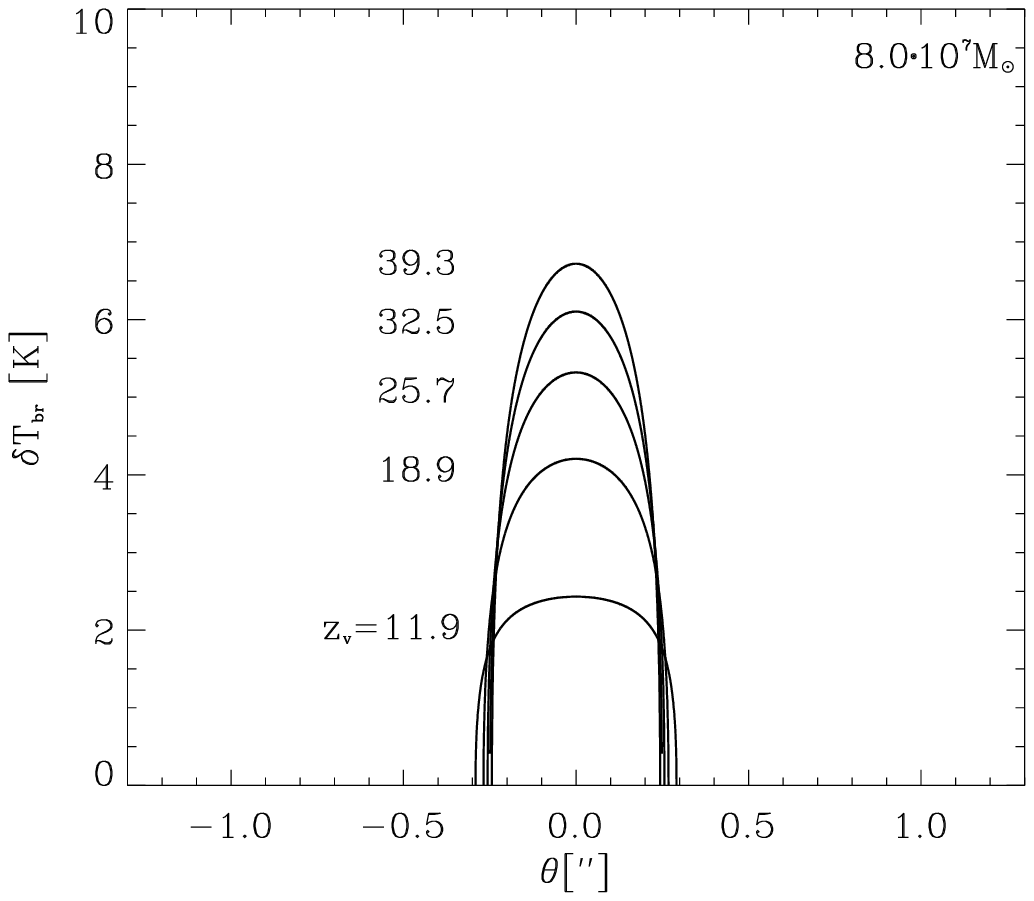}
\caption{The profiles of brightness temperature just after virialization of halos with masses $5.1\cdot10^9$ M$_{\odot}$ (left), $6.4\cdot10^8$ M$_{\odot}$ (central) and $8.0\cdot10^7$ M$_{\odot}$ (right), and initial amplitudes of curvature perturbations $C_k=3.0\cdot10^{-4}$, $2.5\cdot10^{-4}$, $2.0\cdot10^{-4}$, $1.5\cdot10^{-4}$, $1.1\cdot10^{-4}$ (from top to bottom). The redshifts of their virialization are indicated.}
\label{profiles}
\end{figure*}
In Figure \ref{profiles} we show the profiles of brightness temperature just after virialization of halos with masses $5.1\cdot10^9$ M$_{\odot}$ (left panel),  $6.4\cdot10^8$ M$_{\odot}$ (central) and $8.0\cdot10^7$ M$_{\odot}$ (right) at different redshifts. Since the observable differential brightness temperature is not homogeneous on the surface of halo we can average it like 
$$\langle\delta T_{br}\rangle=2\int_0^1\delta T_{br}(x)xdx,$$
where $x=\theta/\theta_h$. In our case the integral has exact analytical presentation
\begin{eqnarray}
\langle\delta T_{br}\rangle=\frac{T_s-T_{CMB}}{1+z}\left[1-\frac{2}{(\tau_{\nu}^0)^2}
+\frac{2e^{-\tau_{\nu}^0}}{\tau_{\nu}^0}+\frac{2e^{-\tau_{\nu}^0}}{(\tau_{\nu}^0)^2}\right].
\label{e_Tbrm}
\end{eqnarray}
\begin{figure*}
\includegraphics[width=0.49\textwidth]{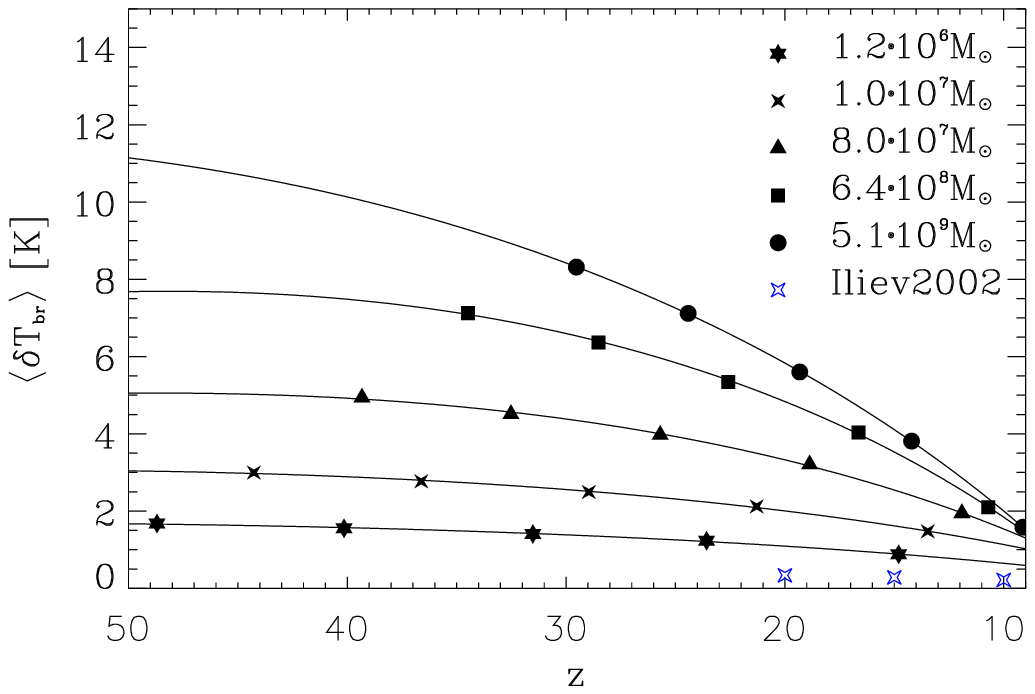}
\includegraphics[width=0.49\textwidth]{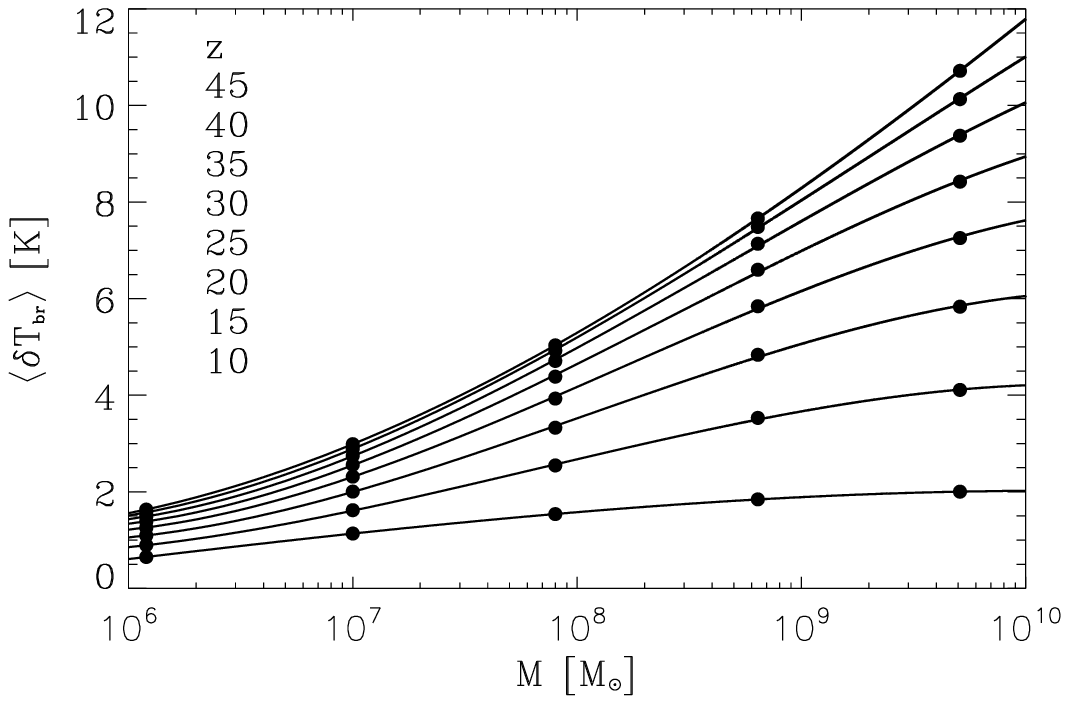}
\caption{The surface averaged brightness temperature of halos with different mass virialized at different $z$. The blue open 4-fold stars show the brightness temperatures of single halos with mass $10^7$ M$_{\odot}$ at $z$=10, 15, 20 from \cite{Iliev2002}.}
\label{TbrzM}
\end{figure*}

It has obvious asymptotic behaviour: if $\tau_{\nu}^0\rightarrow0$ then $\langle\delta T_{br}\rangle\rightarrow0$, if $\tau_{\nu}^0\rightarrow\infty$ then $\langle\delta T_{br}\rangle\rightarrow (T_s-T_{CMB})/(1+z)$. 
The averaged differential brightness temperatures $\langle\delta T_{br}\rangle$ for different dark ages halos are presented in Table \ref{Tab2} and \ref{Tab2A}, and in Figure \ref{TbrzM}. They are larger for halos which were formed earlier since the spin temperatures in them are higher. The lower the halo mass, the less is its opacity, and, consequently, the greater the effect of ``darkening to the edge of the disc.'' The dependence of averaged brightness temperature on redshift for halo with fixed mass $M_h$ is approximated well by simple formula 
$\langle\delta T_{br}\rangle=\alpha+\beta z+\gamma z^{1.2}$, which is shown by solid lines in the left panel. The best-fit coefficients ($\alpha,\,\beta,\,\gamma$) are presented in Appendix A (Table \ref{TabTbz}). The dependence of averaged brightness temperature on halo mass $M_h$ at fixed redshifts is approximated well by simple formula $\langle\delta T_{br}\rangle=a+b\cdot\lg M_h+c\cdot\lg^2M_h+d\cdot\lg^3M_h$, which is shown by solid lines in the right panel. The best-fit coefficients ($a,\,b,\,c,\,d$) are presented in Appendix A (Table \ref{TabTbM}). 

The dependence of differential brightness temperature of individual halos on their mass at different redshift has been computed by \cite{Iliev2002} (Fig.1). For the halos with mass of $10^7$ M$_\odot$ at $z$=10, 15 and 20 it gives $\delta T_{br}\approx$0.22, 0.29 and 0.34 accordingly. These values are shown in the left panel of Figure \ref{TbrzM} by blue open 4-fold stars. They are lower than corresponding values on the line with dark 4-fold filled stars by $\sim$5-6 times. The difference is caused by the dissimilarity in models of sources structure resulted in the different opacity and gas kinetic temperature.  Really, when we recalculate their differential brightness temperatures to the opacity of our halos, then these values increase by 10-20 times and will be above our line for halo the same mass, since the gas kinetic temperatures are higher in those models. 

Thus, the atomic hydrogen in halos is a source of emission in the 21 cm line because the gas there is warm and the kinetic temperature is essentially higher than CMB one. But surrounding gas is cold, its kinetic temperature is lower than CMB one and it can absorb the photons emitted by halos as it was noted by \cite{Furlanetto2006}. In our case when the line is narrow and IHM expands according to the Hubble-Lema\^{\i}tre law the optical depth of absorbing layer is as follows: $\tau_{IHM}(z)=c\alpha_{\nu}^{IHM}(\Delta\nu/\nu)_h/H(z)$. The estimation shows that its value is in the range of 0.01-0.03 at $z$=10-50. It means that surrounding gas absorbs only about $\sim$1-3\% of halo radiation. The absorption can be evaluated more accurate when the inhomogeneity of gas distribution around halo is taken into account. Authors of \cite{Xu2018} have shown that such absorption in some cases can reach dozens percents. But, since, it does not change the predicted brightness temperature of halos dramatically, we omit the similar effects in this paper.

\begin{figure*}
\includegraphics[width=0.32\textwidth]{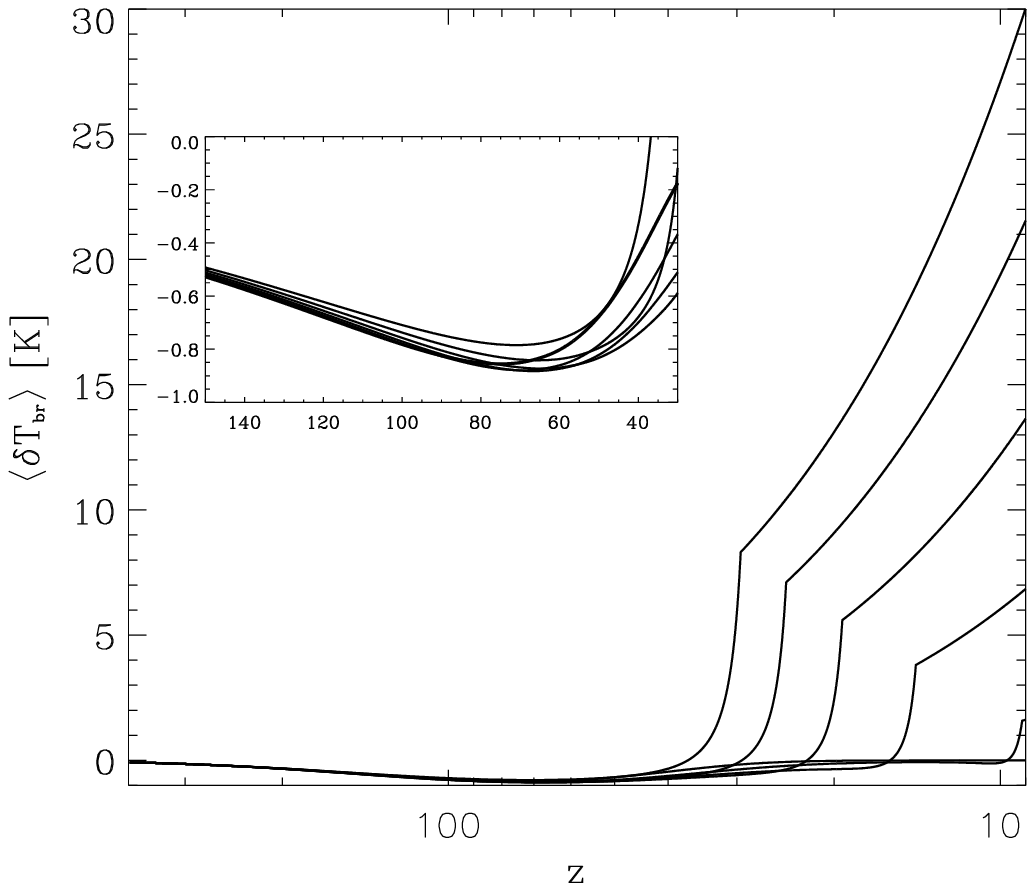}
\includegraphics[width=0.32\textwidth]{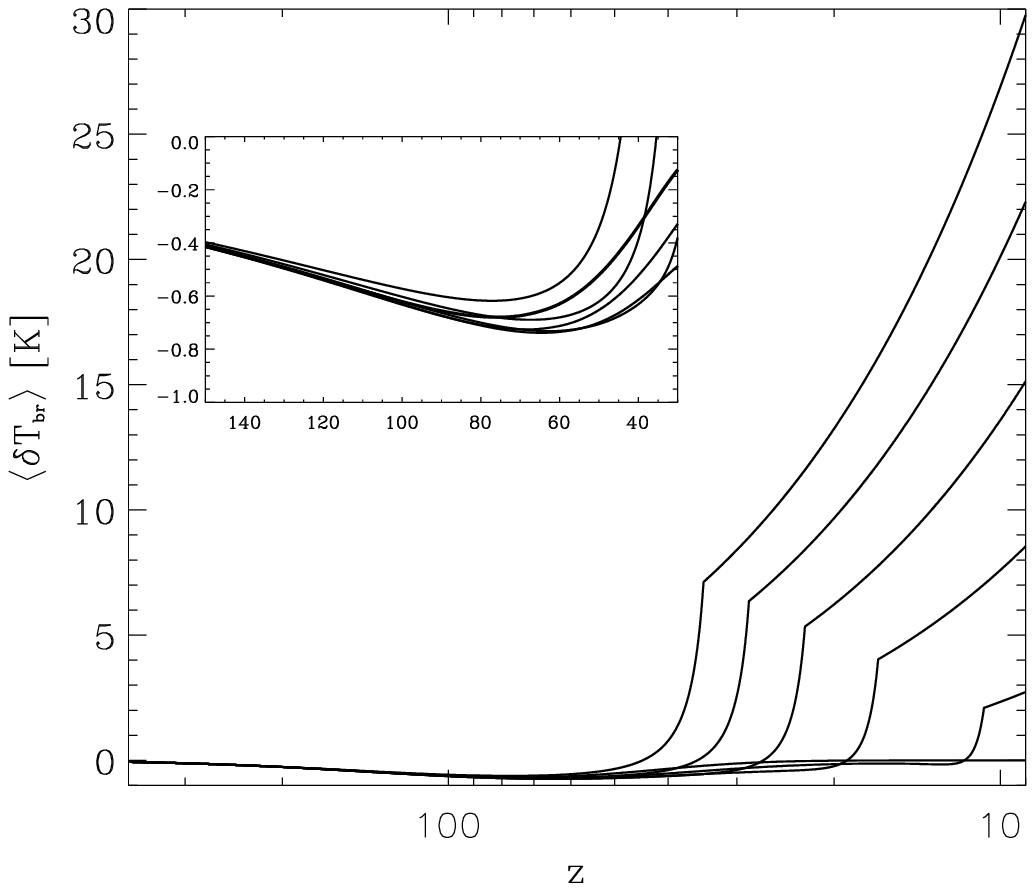}
\includegraphics[width=0.32\textwidth]{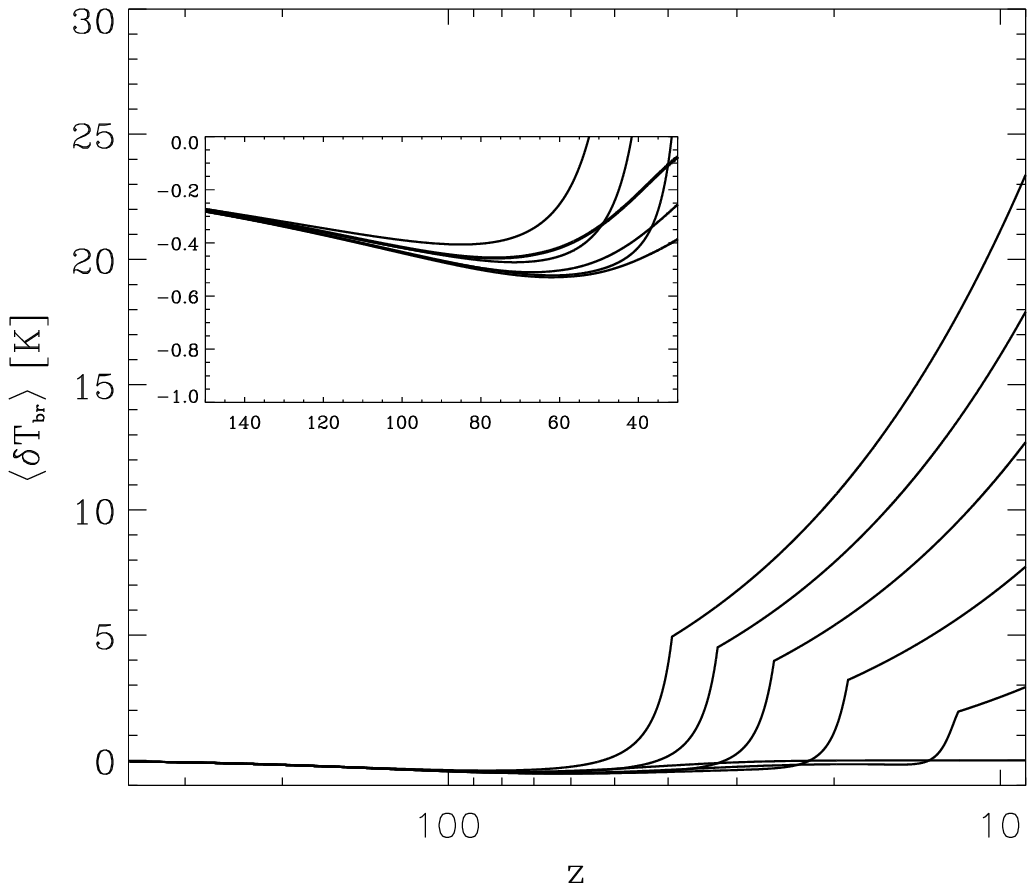}
\caption{The evolution of surface-averaged brightness temperatures of individual halos with masses $5.1\cdot10^9$ M$_{\odot}$ (left), $6.4\cdot10^8$ M$_{\odot}$ (central) and $8.0\cdot10^7$ M$_{\odot}$ (right), and decreasing initial amplitudes of curvature perturbations 
$C_k=3.0\cdot10^{-4}$, $2.5\cdot10^{-4}$, $2.0\cdot10^{-4}$, $1.5\cdot10^{-4}$, $1.1\cdot10^{-4}$ and $C=0$ (from top to down). }
\label{Tbr_evol}
\end{figure*}

In our approach we can look also how the differential brightness temperature of single halo changes before and after virialization. Its evolution for halos virialized at different redshifts $z_v$ is shown in Figure \ref{Tbr_evol}. In the inserts, it is shown a period before the turnaround, when the corresponding halo could be observed in the  absorption of the 21 cm line. Before of turnaround and virialization the changes of differential brightness temperature is caused by the changes of spin temperature and opacity mainly, after virialization by decreasing of CMB temperature and increasing of frequency with decreasing of redshift.

\section{The antenna temperature}

Let us estimate the beam-averaged brightness temperature of the dark ages halos which are in the field of view of a radio telescope.  

For observations of 21 cm ($\nu_0=1420$ MHz) emission from halos placed at the redshift $z$ the radio telescope must be tuned on the redshifted frequency $\nu_{obs}=\nu_0/(1+z)$ and the frequency range $\Delta\nu_{fr}$ of integration of signals. The halos from the redshift range $[z,\,\, z+\Delta z]$ will be registered in the frequency range [$\nu_{obs},\,\nu_{obs}-\Delta\nu_{fr}$]\footnote{We suppose that the thermal broadening of the line and its shift caused by peculiar velocity of halo are much lower than $\Delta\nu_{fr}$.}. The redshift and frequency range are related as 
\begin{equation}
\Delta z =(1+z)^2\frac{\Delta\nu_{fr}}{\nu_{obs}}\left(1-\frac{\Delta\nu_{fr}}{\nu_{obs}}\right)^{-1}\nonumber.
\end{equation}
It means that the surface number density of halos which a radio telescope can detect in the frequency range [$\nu_{obs},\,\,\nu_{obs}-\Delta\nu_{fr}$] is  as follows 
\begin{equation}
\sigma_h(M,z)=\frac{c}{H_0}\int_z^{z+\Delta z}n_h(M\ge M_{min};z)\frac{dz'}{E(z')}\,\,\rm{Mpc}^{-2},
\label{sigmah1}
\end{equation}
where 
$$n_h(M\ge M_{min};z)=\int_{M_{min}}^{\infty}\frac{dn}{dM}dM,$$
is the number density of halos with masses $M\ge M_{min}$, $E(z)\equiv\sqrt{\Omega_m(z+1)^3+\Omega_{de}(z+1)^{3(1+w_{de})}}$.\,\, In\, the\, case $\Delta\nu_{fr}\ll\nu_{obs}$ it can be simplified as follows  
\begin{eqnarray}
&&\hskip-0.7cm \sigma_h(M,z)=\nonumber\\
&&\quad 5.25 \left(\frac{\Delta\nu_{fr}}{1\,\rm{MHz}}\right)\frac{n_h(M,z)}{E(z)}\left[\int_0^z\frac{dz'}{E(z')}\right]^2\, {\rm arcmin^{-2}}. \nonumber
\end{eqnarray}

We present a surface number density of halos in units of arcmin$^{-2}$ in Table \ref{Tab2} and \ref{Tab2A} for different redshifts. 
The most of radio telescopes working in the frequency range $\sim100$ MHz have beam widths from few arcminutes to few degrees, essentially larger than the angular size of halos. The mean angular distance between halos virialized at $10\le z\le 30$ is
$\theta_{mad}(z)\approx\sqrt{\sigma^{-1}_h(z)}\sim0.3'-10^o$, which depends on the mass of halo and redshift of its virialization. The number of halos of mass $M_h$ virialized at $z,\,z+\Delta z$ and which are in the field of view of antenna with beam diameter $\theta_{beam}$ is as follows
$$N_{beam}(M_h\ge M_{min};z)=\frac{\pi}{4}\theta^2_{beam}\sigma_h(M_h\ge M_{min};z).$$
For example, in the field of view of an antenna with $10'$ beam width in the frequency band 1 MHz centered at 130 MHz are $\sim10^4$ halos with $M_h>10^6$ M$_{\odot}$, in the frequency band 1 MHz centered at 90 MHz are $\sim800$ halos, centered at 70 MHz are $\sim30$ halos and so on.

\begin{figure*}
\includegraphics[width=0.49\textwidth]{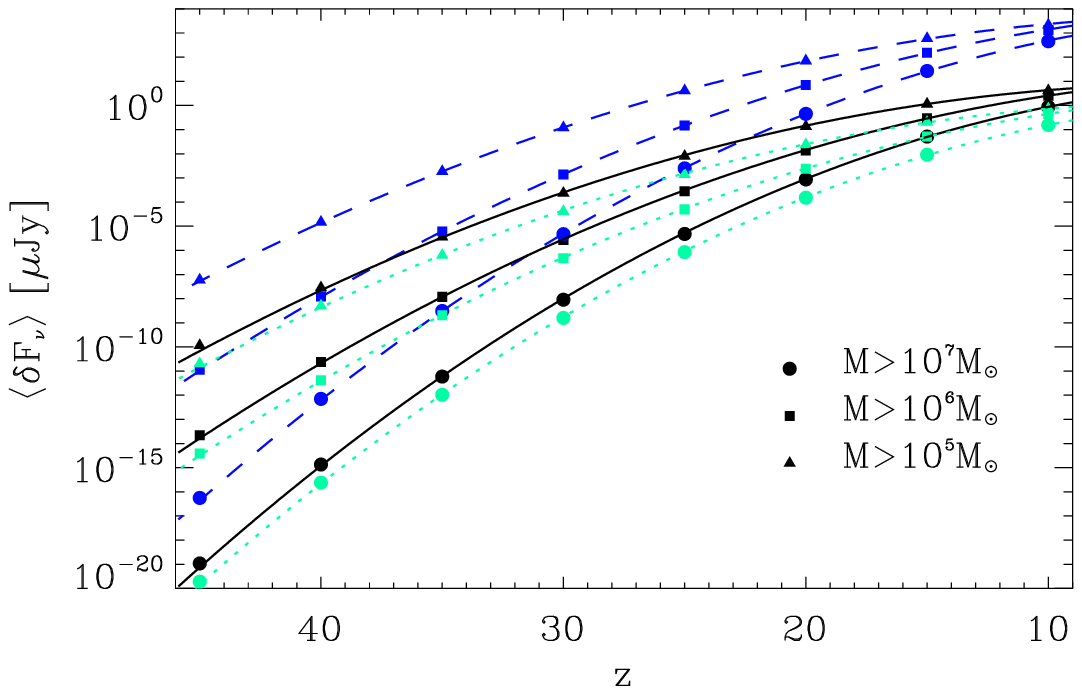}
\includegraphics[width=0.49\textwidth]{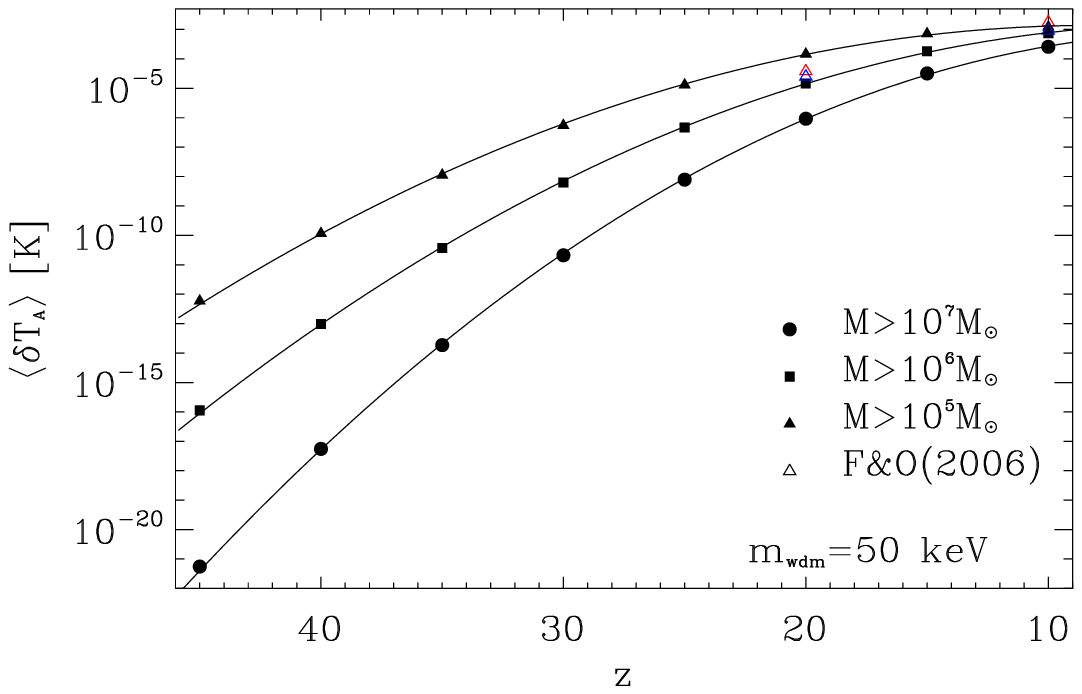}
\caption{The beam-averaged differential flux per unit frequency $\widehat{\delta F}_{\nu}$ (left panel) and beam-averaged effective differential antenna temperature $\widehat{\delta T}_A$ (right panel) from all halos observed within a given antenna beam of angular size $\theta_{beam}=3.8^o$ (blue dashed lines), 10' (dark solid lines) and 4.2' (green dotted lines) in the frequency band $\nu_{obs}\, -\,\nu_{obs}+\Delta\nu_{eff}$. The results are shown for the halos with $M_h\ge10^5$ M$_\odot$, $M_h\ge10^6$ M$_\odot$ and $M_h\ge10^7$ M$_\odot$. The mass functions at different $z$ correspond to  $\Lambda$WDM model with $m_{WDM}>50$ keV, shown by solid lines in Figure \ref{hmf}. The open triangles show the expected rms brightness temperature fluctuations from minihalos at $k=5$ Mpc$^{-1}$ with gas kinetic temperature 1000 K (red) and 100 K (blue) from \cite{Furlanetto2006} .}
\label{adtf}
\end{figure*}

\begin{figure*}
\includegraphics[width=0.49\textwidth]{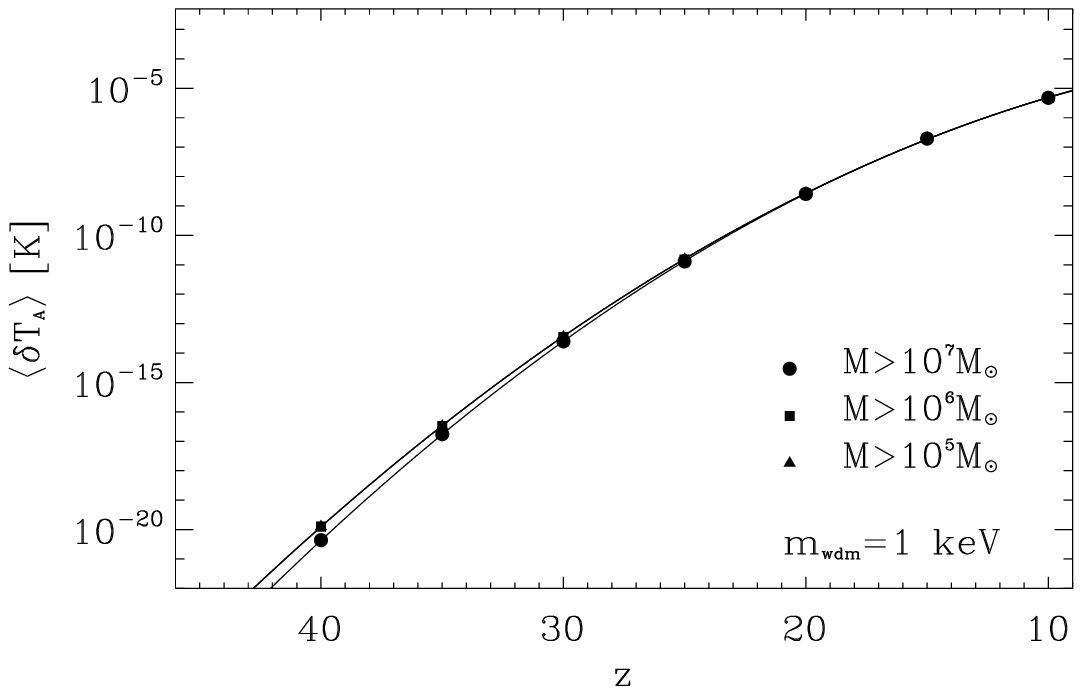}
\includegraphics[width=0.49\textwidth]{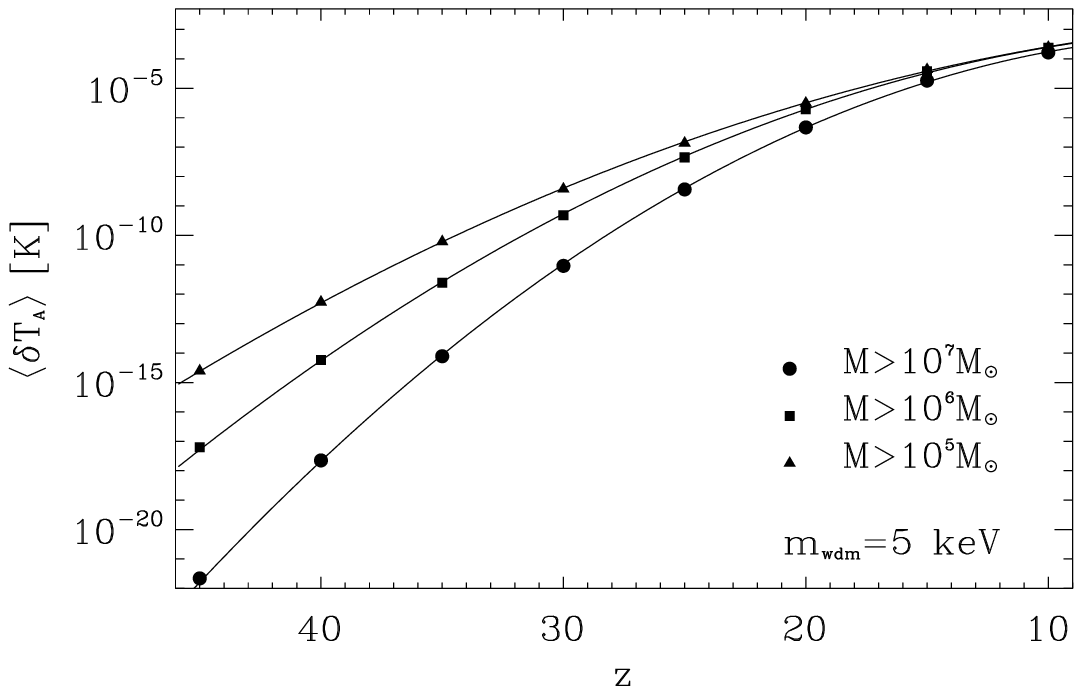}
\caption{The beam-averaged effective differential antenna temperature $\widehat{\delta T}_A$ from halos with $M_h\ge10^5$ M$_\odot$, $M_h\ge10^6$ M$_\odot$ and $M_h\ge10^7$ M$_\odot$ for WDM with $m_{WDM}=$ 1 and 5 keV. The mass functions at different $z$ correspond to $\Lambda$WDM model are shown by dotted lines in Figure \ref{hmf}.}
\label{adt1-5}
\end{figure*}

When the differential brightness temperatures and angular radius of individual halo are known, we can compute the differential energy flux per unit frequency as
\begin{eqnarray}
&&\hskip-1.7cm\left(\frac{d\delta F}{d\nu}\right)_{obs}=2\pi\left(\frac{\nu_{obs}}{c}\right)^2k_B\langle\delta T_{br}\rangle\theta_h^2\\
&&= 0.227\left(\frac{\nu_{obs}}{100\,\rm{MHz}}\right)^2\left(\frac{\langle\delta T_{br}\rangle}{10\,\rm{K}}\right)\left(\frac{\theta_h}{1''}\right)^2\,\,\rm{\mu Jy}\nonumber
\end{eqnarray}   
The values of differential energy fluxes $\left(d\delta F/d\nu\right)_{obs}$ for the analysed halos are presented in Table \ref{Tab2} and \ref{Tab2A}. Taking into account the surface number density of halos of different mass one conclude that the range of frequencies for search of the halo thermal emission signal in the redshifted 21 cm line of atomic hydrogen by radio telescopes is 70-130 MHz. 

The average differential flux per unit frequency from all halos observed within a given antenna beam of angular size $\theta_{beam}$ in the frequency band $\nu_{obs}\, -\,\nu_{obs}+\Delta\nu_{obs}$ can be estimated according to the formula followed from \cite{Iliev2002}, 
\begin{equation}
\widehat{\delta F}_{\nu}(z)=\frac{\pi}{2}\frac{ck_B\nu_0}{H_0^3}\frac{1+z}{E(z)}\theta^2_{beam}\mathcal{I}(z;M_h) ,
\end{equation}
where
$$\mathcal{I}(z;M_h)\equiv\pi\left[\int_0^z\frac{dz}{E(z)}\right]^2\int_{M_h}^{\infty}\langle\delta T_{br}\rangle\Delta\nu_{eff}\theta_h^2\frac{dn}{dM}dM.$$
The effective observable width of the line is caused by thermal Doppler broadening (\ref{Dnuz}) divided by $(1+z)$. Since the gas temperature in our model is defined by the redshift of virialization the effective width is well approximated by simple relation: $\Delta\nu_{eff}=798.33-4.6429z$.  The angular sizes of beam and halo, $\theta_{beam}$ and $\theta_h$, are in radians here. Each halo also has peculiar velocity caused by cosmological perturbations of larger scales. The rms value can be estimated from the initial power spectrum of density perturbations as 
$\langle V^2_p(M_h,z)\rangle^{1/2}=H(z)\left[\int_0^{\infty}P(k,z)W^2(kr_h)dk/2\pi^2\right]^{1/2}$. Such motion shifts the line of individual halo for $\pm\sim 0.1$ MHz that practically does not influence our estimations. 

The results are presented as dependences of spectral fluxes on redshift for three mass ranges, $M_h\ge 10^7$ M$_\odot$,  $M_h\ge 10^6$ M$_\odot$ and $M_h\ge 10^5$ M$_\odot$ in Figure \ref{adtf} (left panel). The mass functions at different $z$ correspond to $\Lambda$WDM model with $m_{WDM}>50$ keV, shown by solid lines in Figure \ref{hmf}. Since the differential spectral flux from halos is proportional to the square beam size of the antenna, we present the estimations for the antenna with the radius of beam $\theta_{beam}=1.9^o$, 10 arcmin and 2.1 arcmin. The first is the radius of LOFAR antenna beam \cite{Gehlot2018}, the last is MWA beam radius \cite{Ewall-Wice2016}. One can see that average differential flux per unit frequency from all halos observed within a given antenna beam fast decreases with increasing of $z$. This is caused by the fast decreasing of the number density of virialized halos with increasing of $z$, as it follows from mass function in Figure \ref{hmf}. The lower is the minimal mass of halos the larger is average differential flux at different redshifts.  

At last, we can estimate the value of beam-averaged effective differential antenna temperature \cite{Iliev2002} as follows 
\begin{equation}
\widehat{\delta T}_A(z,M_h)=\pi\left(\frac{c}{H_0}\right)^3\frac{(1+z)^3}{\nu_0E(z)}\mathcal{I}(z;M_h). 
\label{idTaz} 
\end{equation}
The results for the same halos are presented in the right panel of Figure \ref{adtf}. They do not depend on the beam radius. One can see, that the beam-averaged antenna temperature of halos decreases with increasing of $z$ and depends on the mass range of observed luminous halos which is related with the initial power spectrum of density perturbations. To analyse how the beam-averaged effective differential antenna temperature depends on the free-streaming scale (or cut-off one) we compute it for $\Lambda$WDM model with $m_{WDM}=$ 1 and 5 keV (Figure \ref{adt1-5}). For the first $R_{fs}\simeq7\cdot10^6$ M$_{\odot}$, for the second $R_{fs}\simeq10^3$ M$_{\odot}$. The lower is the mass of WDM particles the large is cut-off of mass function, the lower is the beam-averaged effective differential antenna temperature of dark ages halos. Really, in the $\Lambda$WDM model with $m_{WDM}\ge50$ keV the expected beam-averaged differential antenna temperature of dark ages halos at $z\sim10$ is order of few millikelvins, for the model with $m_{WDM}=5$ keV it is $\sim0.1$ millikelvins, with $m_{WDM}=1$ keV it is $\sim0.01$ millikelvins. The lower is $m_{WDM}$ (the large cut-off) the faster decreases the beam-averaged temperature with increasing of redshift. So, the tomography of the Dark Ages in 21 cm line of atomic hydrogen will give us the information about short wave cut-off of initial power spectrum which can shed the light on the type of dark matter, cold or warm.

Note, that these results also need to be corrected for the beam efficiency coefficient $\eta_{be}$ to obtain the observable antenna temperature. For the most antenna it lowers the amplitude by factor $\sim0.9-0.5$.

\section{Discussions and comparisons}

The expressions (\ref{Tsa}) and (\ref{e_dTbr}) state that dark ages halos are sources of thermal emission in the 21 cm hyperfine line of atomic hydrogen always when their temperature $T_K$ is higher than $T_{CMB}$ at the same redshift. Really, these two expressions can be reduced to one,
\begin{equation}
\delta T_{br}=2.73\frac{T_0}{T_{CMB}}\frac{T_K-T_{CMB}}{T_K+T_0}(1-e^{-\tau_{\nu}}).
\label{dTbr-TK}
\end{equation}
Our analysis shows that halos which are virialized at $z\ge 10$ have kinetic temperature essentially higher than the temperature of CMB. 
We assumed that virialized halos have an adiabatic temperature which they achieve at the beginning of virialization when density contrast becomes $\sim178$. 
In the hierarchical scenario of halo formation it is usually accepted that practically all difference of gravitational potential energy of halo baryonic mass in turn-around point and after violent virialization goes into heating of the gas. Then according to the virial theorem the temperature of the gas in the halo with mass $M_h$ virialized at the redshift range $z\sim10-50$ will be in the range \cite{Barkana2001,Bromm2011} 
$$T_v=(2-9)\cdot10^4\left(\frac{M_h}{10^8\,{\rm M}_\odot}\right) \quad {\rm K}.$$
And only halos with mass $M_h\le10^7-10^8$ M$_\odot$ have $T_v<10^4$ K which provide enough neutral hydrogen to produce the 21-cm signal. It is so in the $\Lambda$CDM model. In the $\Lambda$WDM model with mass of halos in the range of $M_{fs}\lesssim M_h\lesssim M_{hm}$ the situation is not well defined, since formation of halos is delayed and hierarchy may fail \cite{Schneider2012}. That is why we carry out all computations  for halos with adiabatic kinetic temperature which is the minimal temperature of the gas, since processes during virialization can only increase it. We believe that for a considerable time, much of the baryon gas in halos has a temperature that is determined by adiabatic compression. Meanwhile, the results can be generalised for other mechanisms of heating in the conditions of Dark Ages when collisional excitation is dominant and fractions are the same.
\begin{figure}
\includegraphics[width=0.49\textwidth]{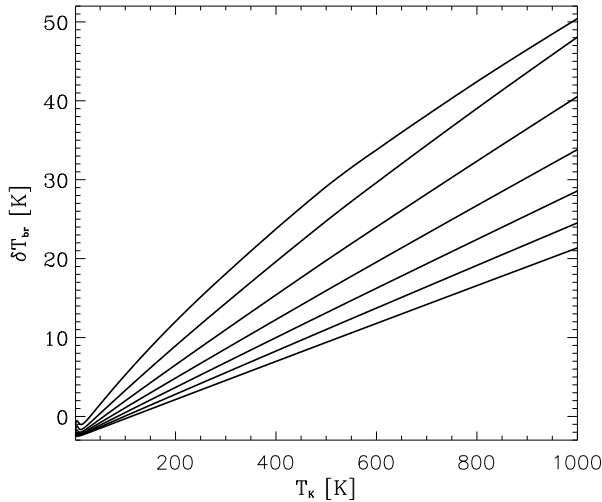}
\caption{The dependence of 21-cm line brightness temperature of individual halos on the kinetic temperature at different redshifts $z=10$, 15, 20, 25, 30, 35, 40 (from top to down).}   
\label{Tbr_T}
\end{figure}
The dependence of 21-cm line brightness temperature of individual halos on the kinetic temperature at different redshifts $z=10-40$ is shown in Figure \ref{Tbr_T} for optically thick halos ($\tau_\nu\gg1$). For the opposite case, optically thin halos ($\tau_\nu\ll1$), the values in the figure must be multiplied by $\tau_\nu$.

The values for the brightness temperature of halos that were retrieved from our model are compared to results obtained in the well-known studies \cite{Iliev2002,Furlanetto2006}. Figure \ref{TbrzM} compiles the results of this paper and those taken from Iliev's one \cite{Iliev2002} (Figure 1). 
Note the main trend: the increase in the brightness temperature of a single halo with an increase in its mass and the redshift of its virialization.  Some quantitative differences in brightness temperatures for a halo with the same mass at the same redshifts are resulted by different models of halos. In our case, the halo opacity increases with the mass of halo and is $\sim1-10$, while in Iliev's model it decreases and is $\ll1$ for halos with $M\sim10^6-10^8$ M$_\odot$. The kinetic temperatures of gas in halos and the width of line from individual halos are different too. Naturally, the beam-averaged fluxes and brightness temperatures presented in Figure \ref{adtf} are some lower than ones in Figure 2 of \cite{Iliev2002} because we do not take into account the halos of lower mass the surface number density of which is much higher. In the right panel of Figure \ref{adtf} we compare the expected beam averaged brightness temperature and rms brightness temperature fluctuations from halos at $k=5$ Mpc$^{-1}$ predicted by Furlanetto \& Oh (2006) for the case when kinetic temperature of gas in halos equals 1000 K and 100 K (Figure 1 in \cite{Furlanetto2006}). Our results for halos with $M_h\ge10^5-10^6$ M$_{\odot}$ are very close to the Furlanetto \& Oh's ones for similar halo model. The ``similar'' here means that hyperfine level is excited by collisions and CMB radiation without Ly-$\alpha$ coupling. But there is a noticeable difference in general parameters of models (density profiles, mass range of minihalos, gas kinetic temperatures etc.) and the comparison of models can be inadequate since in the $\Lambda$CDM and $\Lambda$WDM models the scenarios of halo formation are different at different scales. But we can compare the expected signal for the different models of halos, for the different approaches of signal estimation etc. We note, that rms brightness temperature fluctuations is statistical measure while the beam-averaged values are related to the individual measurements. Bringing them to the same angular or spatial scale is not an obvious and simple step in a such comparison. Nevertheless, we can state the proximity of values for congruent scales the with the same excitation mechanisms. Since the power spectrum of brightness temperature fluctuations from halos is defined via integral of the mass function of halos (see eq. (6) in \cite{Furlanetto2006}) like the beam averaged brightness temperature in this paper (eq. \ref{idTaz}), then one can assume that the dependence of power spectrum on the mass of WDM particles (or collisionless cut-off) will be similar as in Figure \ref{adt1-5}.

Let's evaluate the possibility of observing the individual halos which have angular sizes 0.01-2 arcsec and brightness temperature 1-50 K in the frequency range 140-30 MHz using existing advanced telescopes. The maximal angular resolution of the LOFAR Low Band Antenna (LBA) array in the frequency range 90-30 MHz is $\sim$15 arcsec \cite{Gehlot2018}. The maximal resolution of the LOFAR High Band Antenna (HBA) array in the frequency range 115-190 MHz is $\sim$6 arcsec \cite{Patil2017}. In \cite{vanHaarlem2013} it is stated that maximum angular resolution can be achieved up to sub arcseconds when all far stations will observe in the interferometric regime. The maximal sensitivity to the redshifted 21-cm signal at the frequency range 50-70 MHz have been achieved by LOFAR LBA at the level of $\sim0.1$ K for 14 hours of time integration \cite{Gehlot2018}. It looks quite optimistically, but the problem is that maximum sensitivity and maximum angular resolution which are necessary cannot be achieved simultaneously. 
Five-hundred-meter Aperture Spherical radio Telescope\footnote{http://fast.ac.cn/} (FAST) has the angular resolution about 3 arcminutes in the frequency range 70 MHz - 3 GHz. The MWA radio telescope (70-300 MHz) has angular resolution a few arcminutes. Others low-frequency telescopes have worse angular resolutions. The future telescopes planned for this frequency range do not seem to solve the problem of observing the individual halos in the Dark Ages. So, the Square Kilometre Array\footnote{https://www.skatelescope.org/}, which will start to observe the sky in the middle of 20s, will provide frequency coverage from 50 MHz to 30 GHz with sensitivity in 50 and more times better than LOFAR. But the maximal angular resolution at the nominal frequency $\sim100$ MHz is planned of 11 arcsec. So, only the beam-averaged signal or the power spectrum of brightness temperature fluctuations caused by dark ages halo statistics can be measured.

Now we compare the obtained estimation of thermal emission of dark ages halos in 21 cm hyperfine hydrogen line with upper limits on the brightness temperature power spectrum which are established in a few experiments and collected in Table \ref{Tab4}. 
\begin{table} 
\begin{center}
\caption{The upper limits on the brightness temperature power spectrum of the redshifted 21-cm signal provided by several instruments. In \cite{Gehlot2018} the LOFAR LBA results are presented for two celestial fields: 3C220.3 galaxy vicinity and North Celestial Pole (in before last and last rows accordingly).}
\begin{tabular} {cccc}
\hline
\hline
   \noalign{\smallskip}
Instrument&$k$&$z$-range&2$\sigma$ upper limit \\
 \noalign{\smallskip}
&$h\,$Mpc$^{-1}$&&(mK)$^2$ \\
 \noalign{\smallskip} 
\hline
   \noalign{\smallskip} 
GMRT \cite{Paciga2013}&0.5&8.6&248 \\
MWA \cite{Beardsley2016}&0.27&7.1&164 \\ 
MWA \cite{Ewall-Wice2016}&0.27&[12, 18]&10$^4$ \\
LOFAR HBA \cite{Patil2017}&0.053&[9.6, 10.6]&79.6 \\
LOFAR LBA \cite{Gehlot2018}&0.038&[19.8, 25.2]&14561 \\
LOFAR LBA \cite{Gehlot2018}&0.038&[19.8, 25.2]&14886 \\
  \hline
\end{tabular}
\label{Tab4}
\end{center}
\end{table}
Here we compare our predictions with upper limits for the highest redshifts given by LOFAR and MWA teams, since for comparison with measurements at lower redshifts  other sources of heating, ionising, excitation and de-excitation (Wouthuysen-Field effect \cite{Wouthuysen1952,Field1958}) must be taken into account for the computation of populations of the hyperfine levels of atomic hydrogen.  

The main parameters of LOFAR observations, which are important in our analysis, are as follows: field of view is 12 deg$^2$ (beam diameter 3.8 deg), the frequency range is $54\div68$ MHz, the redshift range is $z=19.8\div25.2$, the integration time is 14 hours and the frequency resolution is 3 kHz \cite{Gehlot2018}. 

Using the mass function of halos and analytical approximations for $\langle\delta T_{br}\rangle(M;z)$, $\Delta\nu_{eff}(z)$ and $\theta_h(M)$ presented above, one can estimate the total number of halos with masses $M_h\ge10^6$ M$_\odot$ ($M_h\ge10^5$ M$_\odot$) in the LOFAR field of view: $N_h^{(tot)}\sim3\cdot10^5$ ($\sim10^{7}$). These halos provide the average differential flux 8.2 (78.7 $\mu$Jy) and the beam-averaged effective differential antenna temperature $\delta T_A=0.017$ (0.16) mK, that is essentially lower than the upper limit established recently by LOFAR \cite{Gehlot2018}. 

Important parameters of MWA observations are: field of view is 14 arcmin$^2$ (beam diameter 4.2 arcmins), the frequency range is $75\div113$ MHz (two bands $75.52\div90.88$ MHz and $98.84\div112.64$ MHz), the redshift range is $z=11.6\div17.9$, the integration time is 3.08 hours and frequency resolution is 40 kHz \cite{Ewall-Wice2016}. In the frequency band $98.84\div112.64$ MHz ($z=11.6\div13.6$), called Band 2, in the field of view of MWA there are about $1.2\cdot10^4$  ($2.1\cdot10^5$) halos of mass $M_h\ge10^6$ M$_\odot$ ($M_h\ge10^5$ M$_\odot$) which provide the average differential flux 0.22 (0.40 $\mu$Jy) and the beam-averaged effective differential antenna temperature $\delta T_A=0.48$ (0.88) mK. In the Band 1, $75.52\div90.88$ MHz, the MWA can see about $\sim3\cdot10^3$ ($\sim7\cdot10^4$) halos which ensure the average differential flux 0.065 (0.24 $\mu$Jy) and the beam-averaged effective differential antenna temperature $\delta T_A=0.22$ (0.81) mK. Both estimations are essentially lower than the upper limit established by \cite{Ewall-Wice2016}. 

The integration time to detect the hyperfine 21 cm emission of the neutral hydrogen with recent low frequency telescopes LOFAR or MWA is estimated as low realistic. But with more sensitive telescopes of the next generation such as SKA or HERA\footnote{https://www.ska.ac.za/science-engineering/hera/}, the  integration times  needed are quite acceptable.

\section{Conclusions}

We analysed the thermal emission of dark ages halos in 21 cm line of atomic hydrogen excited by collisions with neutral and ionised hydrogen and electrons. We suppose the halos are formed in the Gaussian density peaks of cosmological scalar mode perturbations. The semianalytical modelling of formation of individual spherical halos in multi-component models shows that the kinetic temperature of gas in halos virialized at $z\ge10$ is higher than the temperature of cosmic microwave background and it is possible to detect them as arcsecond sources of emission in the frequency range $70-130$ MHz. It is shown that in the dark ages halos the inelastic collisions between neutral hydrogen atoms are the dominant excitation mechanism of hyperfine structure levels, which pulls the spin temperature closer to the kinetic one. We have estimated the brightness temperatures of dark ages halos with masses $10^6-10^{10}$ M$_\odot$ and adiabatic temperatures $\sim60-800$ K which are virialized at $z\sim10-50$. It is shown that their brightness temperatures are in the range 1-10 K, increase with the mass of halo and redshift of its virialization. The differential flux per unit frequency $(d\delta F/d\nu)_{obs}$ from individual halo is low and varies over four order for these ranges of halo masses and redshifts, $\sim10^{-2}-10^{2}$ nJy, that is a challenge for their detection by existing and planned radio telescopes. The maximal flux $\sim0.1\,\mu$Jy is expected from the halo of the largest mass: the faster it virialized, the greater is its brightness temperature. 

The number density of virialized halos exponentially decreases with increasing of halo mass and redshift. Assuming the 1 MHz frequency band of detection we have estimated the surface number density of halos at different redshifts and beam-averaged effective antenna temperatures and fluxes caused by halos of different masses which are in the field of view of a radio telescope. It turned out that the halos of smaller masses cause a larger beam-averaged differential antenna temperature and flux since their surface number density is higher. We have illustrated also that the beam-averaged signal strongly depends on the cutoff scale in the mass function of dark ages halos which can be caused by free-streaming of WDM particles.   

At last, we compare our results with the first power spectrum limits on the 21-cm signal of neutral hydrogen from MWA \cite{Ewall-Wice2016} and LOFAR \cite{Gehlot2018}: they are essentially below these upper limits. Radical increasing the integration time of observations by MWA and LOFAR will inevitably lead to a long-awaited detection of a signal from the Dark Ages, which will be intensively investigated by the next generation telescopes.

\section*{Acknowledgements}
This work was supported by the International Center of Future Science of Jilin University (P.R.China) and the project of Ministry of Education and Science of Ukraine ``Formation and characteristics of elements of the structure of the multi-component Universe, gamma radiation of supernova remnants and observations of variable stars'' (state registration number 0119U001544).  We acknowledge the anonymous referee for his attentive and accurate report, and useful comments and suggestions.

\section*{Appendix A. Tables and approximation coefficients}

We present here the physical values and chemical compositions of halos with $M_h=1.0\cdot10^7$ and $1.2\cdot10^6$ M$_\odot$ which are virialized at different redshifts (Table \ref{Tab1A}), the parameters of these halos as emitters in 21 cm hydrogen line (Table \ref{Tab2A}) and coefficients of approximations of dependences of averaged brightness temperatures of halo on redshift of virialization (Table \ref{TabTbz}) and mass of halo (Table \ref{TabTbM}).
\begin{table*}
\caption{Physical parameters of halos virialized at different redshifts (continue of Table \ref{Tab1}).}
\begin{tabular} {cccccccccc}
\hline
\hline
   \noalign{\smallskip}
$M_h$&$k$&$C_k$&$z_v$&$T_K$&$\rho_{m}$&$n_{HI}$&$n_p\approx n_e$&$r_h$&$\theta_h$\\
 \noalign{\smallskip} 
[M$_{\odot}$]&[Mpc$^{-1}$]&  & &[K] &[g/cm$^3$]&[cm$^{-3}$]&[10$^{-6}$cm$^{-3}$]&[kpc]&['']\\
 \noalign{\smallskip} 
\hline
   \noalign{\smallskip} 
$1.0\cdot10^7$&40&$3.0\cdot10^{-4}$& 44.3&     713.4&  4.40$\cdot10^{-23}$& 3.14 &     260.6&      0.15&      0.12\\
    \noalign{\smallskip}
              & &$2.5\cdot10^{-4}$& 36.6&     541.7&  2.52$\cdot10^{-23}$&  1.80 &     162.6&      0.19&      0.12\\ 
    \noalign{\smallskip}
              & &$2.0\cdot10^{-4}$& 29.0&     381.3&  1.28$\cdot10^{-23}$&  0.91 &      90.2&      0.23&      0.13\\  
    \noalign{\smallskip}
              & &$1.5\cdot10^{-4}$& 21.3&     235.6&  5.25$\cdot10^{-24}$&  0.37 &      41.2&      0.31&      0.13\\  
    \noalign{\smallskip}
              & &$1.0\cdot10^{-4}$& 13.5&     111.2&  1.44$\cdot10^{-24}$&  0.10 &      12.4&      0.48&      0.14\\              
 \noalign{\smallskip}
 \noalign{\smallskip}
$1.2\cdot10^6$&80&$3.0\cdot10^{-4}$& 48.7&     805.3&  5.82$\cdot10^{-23}$&  4.15 &     321.4&      0.07&      0.06\\ 
 \noalign{\smallskip}
              &  &$2.5\cdot10^{-4}$& 40.1&     615.8&  3.30$\cdot10^{-23}$&  2.35 &     197.0&      0.08&      0.06\\ 
 \noalign{\smallskip} 
              &  &$2.0\cdot10^{-4}$& 31.5&     431.3&  1.63$\cdot10^{-23}$&  1.16 &     105.2&      0.11&      0.06\\ 
 \noalign{\smallskip}
              &  &$1.5\cdot10^{-4}$& 23.6&     276.3&  7.04$\cdot10^{-24}$&  0.50 &      52.7&      0.14&      0.06\\ 
 \noalign{\smallskip}
              &  &$1.0\cdot10^{-4}$& 14.8&     125.4&  1.78$\cdot10^{-24}$&  0.13 &      14.5&      0.22&      0.07\\                   
  \hline 
\end{tabular}
\label{Tab1A}
\end{table*}

\begin{table}
\caption{The best-fit coefficients of analytical approximation of dependence of averaged brightness temperature on redshift, $\langle\delta T_{br}\rangle=\alpha+\beta z+\gamma z^{1.2}$, for halos with fixed mass $M_h$. }
\begin{tabular} {cccc}
\hline
\hline
   \noalign{\smallskip}
$M_h$&$\alpha$&$\beta$&$\gamma$\\
 \noalign{\smallskip}
\hline
   \noalign{\smallskip} 
$5.1\cdot10^9{\rm{M_\odot}}$&-4.4186& 1.5136& -0.54983   \\
$6.4\cdot10^8{\rm{M_\odot}}$&-3.7231& 1.4220& -0.54599 \\  
$8.0\cdot10^7{\rm{M_\odot}}$&-1.7774& 0.84534& -0.32413 \\ 
$1.0\cdot10^7{\rm{M_\odot}}$&-0.39285& 0.3744& -0.13979 \\ 
$1.2\cdot10^6{\rm{M_\odot}}$&-0.10855& 0.18202& -0.066946 \\ 
  \noalign{\smallskip}
  \hline
\end{tabular}
\label{TabTbz}
\end{table}
\begin{table}
\caption{The best-fit coefficients of analytical approximation of dependence of averaged brightness temperature on mass of halo, $\langle\delta T_{br}\rangle=a+b\cdot\lg M_h+c\cdot\lg^2M_h+d\cdot\lg^3M_h $ at different redshifts $z$. }
\begin{tabular} {ccccc}
\hline
\hline
   \noalign{\smallskip}
$z$&$a$&$b$&$c$&$d$\\
 \noalign{\smallskip}
\hline
   \noalign{\smallskip} 
10&-1.662& -0.0850& 0.0125& -0.0079673\\
15&23.980& -10.457& 1.4792& -0.063124\\
20&39.231& -16.577& 2.2671& -0.094117\\
25&46.801& -19.550& 2.6355& -0.10723\\
30&48.051& -19.935& 2.6587& -0.10563\\
35&43.873& -18.096& 2.3856& -0.091413 \\
40&34.902& -14.293& 1.8509& -0.066054\\
45&21.619& -8.7261& 1.0812& -0.030692 \\ 
  \noalign{\smallskip}
  \hline
\end{tabular}
\label{TabTbM}
\end{table}
 
\begin{table*}
\begin{center}
\caption{Parameters of halos as emitters of 21 cm line (continue of Table \ref{Tab2}).}
\begin{tabular} {cccccccccc}
\hline
\hline
   \noalign{\smallskip}
$M_h$&$C_k$ &$\nu_{obs}$&$\Delta\nu_{fr}$&$T_s$&$\tau_{\nu}^0$&$\langle\delta T_{br}\rangle$&$\delta\left(\frac{dF}{d\nu}\right)_{obs}$&$\theta_{hwhm}$&$\sigma_h $\\
 \noalign{\smallskip} 
[M$_{\odot}$]& &[MHz]&[kHz]&[K]& &[K]&[$\rm{nJy}$]&[arcsec] &[arcmin$^{-2}$]\\ 
 \noalign{\smallskip} 
\hline
   \noalign{\smallskip} 
$1.0\cdot10^7$ &$3.0\cdot10^{-4}$& 31.4&  0.598&  694.8&    0.41&    3.00&    0.09&    0.11&  8.80$\cdot10^{-19}$\\ 
    \noalign{\smallskip}
               &$2.5\cdot10^{-4}$& 37.8&  0.627&  522.3&    0.44&    2.77&    0.13&    0.11&  9.56$\cdot10^{-13}$\\ 
    \noalign{\smallskip}
               &$2.0\cdot10^{-4}$& 47.4&  0.661&  360.3&    0.48&    2.49&    0.20&    0.11&  9.47$\cdot10^{-8}$\\  
    \noalign{\smallskip}
               &$1.5\cdot10^{-4}$& 63.7&  0.698&  214.1&    0.57&    2.12&    0.33&    0.12&  8.60$\cdot10^{-4}$\\  
    \noalign{\smallskip}
               &$1.0\cdot10^{-4}$& 98.2&  0.739&   92.2&    0.81&    1.48&    0.62&    0.13&  0.769\\         
 \noalign{\smallskip}     
 \noalign{\smallskip} 
$1.2\cdot10^6$ &$3.0\cdot10^{-4}$& 28.6&  0.579&  788.0&    0.21&    1.68&    0.01&    0.05&  3.21$\cdot10^{-16}$\\ 
 \noalign{\smallskip}
               &$2.5\cdot10^{-4}$& 34.5&  0.611&  596.9&    0.21&    1.55&    0.02&    0.05&  1.31$\cdot10^{-10}$\\
 \noalign{\smallskip} 
               &$2.0\cdot10^{-4}$& 43.7&  0.647&  411.0&    0.23&    1.41&    0.02&    0.05&  5.89$\cdot10^{-6}$\\ 
 \noalign{\smallskip}
               &$1.5\cdot10^{-4}$& 57.8&  0.685&  254.8&    0.27&    1.23&    0.04&    0.06&  1.46$\cdot10^{-2}$\\ 
 \noalign{\smallskip}
               &$1.0\cdot10^{-4}$& 89.9&  0.718&  106.2&    0.38&    0.88&    0.08&    0.06&  8.41\\               
  \hline  
\end{tabular}
\label{Tab2A}
\end{center}
\end{table*}


\begin{thebibliography}{}
\bibitem {Pritchard2012} J.R. Pritchard, A. Loeb,  2012, 21 cm cosmology in the 21st century, Rep. Prog. Phys. 75, id. 086901 https://iopscience.iop.org/article/10.1088/0034-4885/75/8/086901
\bibitem {Barkana2001} R. Barkana, A. Loeb, In the beginning: the first sources of light and the reionization of the universe, Phys. Rep. 349 (2001) 125-238 https://linkinghub.elsevier.com/retrieve/pii/S0370157301000199
\bibitem {Fan2006} X. Fan, C.L. Carilli, B. Keating, Observational Constraints on Cosmic Reionization, Annu. Rev. Astron. Astroph. 44 (2006) 415-462 https://doi.org/10.1146/annurev.astro.44.051905.092514
\bibitem {Furlanetto2006a} S.R. Furlanetto, S.P. Oh, F.H. Briggs, Cosmology at low frequencies: The 21 cm transition and the high-redshift Universe, Phys. Rep. 433 (2006) 181-301 https://doi.org/10.1016/j.physrep.2006.08.002
\bibitem {Bromm2011} V. Bromm, N. Yoshida, The First Galaxies, Annu. Rev. Astron. Astroph. 49 (2011) 373-407
https://doi.org/10.1146/annurev-astro-081710-102608
\bibitem {Galli2013} D. Galli, F. Palla, The Dawn of Chemistry, Annu. Rev. Astron. Astroph. 51 (2013) 163-206
https://doi.org/10.1146/annurev-astro-082812-141029
\bibitem {Bowman2018} J.D. Bowman, A.E.E. Rogers, R.A. Monsalve et al., An absorption profile centred at 78 megahertz in the sky-averaged spectrum, Nature 555 (2018) 67-70 https://doi.org/10.1038/nature25792 
\bibitem {Hills2018} R. Hills, G. Kulkarni, P.D. Meerburg, E. Puchwein, Concerns about modelling of the EDGES data, Nature 564 (2018) E32-E34 https://doi.org/10.1038/s41586-018-0796-5
\bibitem {Ewall-Wice2016} A. Ewall-Wice, J.S. Dillon, J.N. Hewitt, A. Loeb, A. Mesinger et al., First limits on the 21 cm power spectrum during the Epoch of X-ray heating, Mon. Not. R. Astron. Soc. 460 (2016) 4320-4347 https://doi.org/10.1093/mnras/stw1022
\bibitem {Gehlot2018} B. K. Gehlot, F. G. Mertens, L. V. E. Koopmans, M. A. Brentjens, S. Zaroubi et al., The first power spectrum limit on the 21-cm signal of neutral hydrogen during the Cosmic Dawn at z = 20-25 from LOFAR,  Mon. Not. R. Astron. Soc. 488 (2019) 4271-4287 https://doi.org/10.1093/mnras/stz1937
\bibitem {Harker2012} G.J.A. Harker, J.R. Pritchard, J.O. Burns, J.D. Bowman, An MCMC approach to extracting the global 21-cm signal during the cosmic dawn from sky-averaged radio observations, Mon. Not. R. Astron. Soc.  419 (2012) 1070-1084 https://doi.org/10.1111/j.1365-2966.2011.19766.x
\bibitem {Novosyadlyj2018} B. Novosyadlyj, V.M. Shulga, W. Han, Yu. Kulinich, M. Tsizh, 2018, Halos in Dark Ages: Formation and Chemistry, Astrophys. J. 865, id. 38 https://doi.org/10.3847/1538-4357/aad7fa
\bibitem {Iliev2002} I.T. Iliev, P.R. Shapiro, A. Ferrara, H. Martel, On the Direct Detectability of the Cosmic Dark Ages: 21 Centimeter Emission from Minihalos, Astrophys. J.  572 (2002) L123-L126 
\bibitem {Iliev2003} I.T. Iliev, E. Scannapieco, H. Martel, P.R. Shapiro, Non-linear clustering during the cosmic Dark Ages and its effect on the 21-cm background from minihaloes, Mon. Not. R. Astron. Soc.  341 (2003) 81-90 https://doi.org/10.1046/j.1365-8711.2003.06410.x
\bibitem {Cooray2002} A. Cooray, R. Sheth, 2002, Halo models of large scale structure, Phys. Rep. 372 (2002) 1-129 https://doi.org/10.1016/S0370-1573(02)00276-4
\bibitem {Furlanetto2006} S.R. Furlanetto, S.P. Oh, Redshifted 21 cm Emission from Minihalos before Reionization, Astrophys. J. 652 (2006) 849-856 https://doi.org/10.1086/508448
\bibitem {Shapiro2006} P.R. Shapiro, K. Ahn, M.A. Alvarez, I.T. Iliev, H. Martel, D. Ryu, The 21 cm Background from the Cosmic Dark Ages: Minihalos and the Intergalactic Medium before Reionization, Astrophys. J. 646 (2006) 681-690 https://doi.org/10.1086/504972
\bibitem {Kuhlen2006} M. Kuhlen, P. Madau, R. Montgomery, The Spin Temperature and 21 cm Brightness of the Intergalactic Medium in the Pre-Reionization era, Astrophys. J. 637 (2006) L1-L4  https://doi.org/10.1086/500548
\bibitem {Planck2018a} Planck Collaboration: Y. Akrami, F. Arroja, M. Ashdown et al., 2018,  Planck 2018 results. I. Overview and the cosmological legacy of Planck https://arxiv.org/abs/1807.06205 
\bibitem {Planck2018b} Planck Collaboration: N. Aghanim, Y. Akrami, M. Ashdown et al. 2018, Planck 2018 results. VI. Cosmological parameters  https://arxiv.org/abs/1807.06209
\bibitem {Leveque1998} R.J. Leveque, D. Mihalas, E.A. Dorfi, E. M\"{u}ller, Computational Methods for Astrophysical Fluid Flow. Springer-Verlag, Berlin, 1998
\bibitem {Seager1999} S. Seager, D.D. Sasselov, D. Scott, A New Calculation of the Recombination Epoch, Astrophys. J. 523 (1999) L1-L5  https://doi.org/10.1086/312250
\bibitem {Seager2000} S. Seager, D.D. Sasselov, D. Scott, How Exactly Did the Universe Become Neutral? Astrophys. J.  Suppl. Ser.128 (2000) 407-430 https://doi.org/10.1086/313388 
\bibitem {Galli1998} D. Galli, E. Palla, The chemistry of the early Universe, Astron. Astrophys. 335 (1998) 403-420 http://articles.adsabs.harvard.edu/pdf/1998A\%26A...335..403G
\bibitem {Novosyadlyj2016} B. Novosyadlyj, M. Tsizh, Yu. Kulinich, 2016, Dynamics of minimally coupled dark energy in spherical halos of dark matter, Gen. Relativ. Grav. 48, id. 30 https://link.springer.com/article/10.1007\%2Fs10714-016-2031-8
\bibitem {Novosyadlyj2017} B. Novosyadlyj, M. Tsizh, Yu. Kulinich, Evolution of density and velocity profiles of dark matter and dark energy in spherical voids, Mon. Not. R. Astron. Soc. , 465 (2017) 482-491 https://doi.org/10.1093/mnras/stw2767
\bibitem {Iliev2001} I.T. Iliev, P.R. Shapiro, The post-collapse equilibrium structure of cosmological haloes in a low-density universe, Mon. Not. R. Astron. Soc. 325 (2001) 468-482  https://doi.org/10.1046/j.1365-8711.2001.04422.x 
\bibitem {Wang2009} J. Wang, S.D.M. White, Are mergers responsible for universal halo properties? Mon. Not. R. Astron. Soc. 396 (2009) 709-717 https://doi.org/10.1111/j.1365-2966.2009.14755.x
\bibitem {Tinker2010} J.L. Tinker, B.E. Robertson, A.V. Kravtsov, A. Klypin, M.S. Warren, G. Yepes, S. Gottl\"{o}ber, The Large-scale Bias of Dark Matter Halos: Numerical Calibration and Model Tests, Astrophys. J. 724 (2010) 878-886 https://iopscience.iop.org/article/10.1088/0004-637X/724/2/878
\bibitem {Smith2011} R.E. Smith, K. Markovic, 2011, Testing the warm dark matter paradigm with large-scale structures, Phys. Rev. D 84, id. 063507 https://doi.org/10.1103/PhysRevD.84.063507
\bibitem {Schneider2013} A. Schneider, R.E. Smith, D. Reed, Halo mass function and the free streaming scale, Mon. Not. R. Astron. Soc. 433 (2013) 1573-1587  https://doi.org/10.1093/mnras/stt829 
\bibitem {Klypin2016} A. Klypin, G. Yepes, S. Gottl\"{o}ber, F. Prada, S. He\ss, MultiDark simulations: the story of dark matter halo concentrations and density profiles, Mon. Not. R. Astron. Soc. 457 (2016) 4340-4359 https://doi.org/10.1093/mnras/stw248 
\bibitem {Press1974} W.H. Press, P. Schechter, Formation of Galaxies and Clusters of Galaxies by Self-Similar Gravitational Condensation, Astrophys. J. 187 (1974) 425-438 http://adsabs.harvard.edu/doi/10.1086/152650
\bibitem {Bond1991} J.R. Bond, S. Cole, G. Efstathiou, N. Kaiser, Excursion Set Mass Functions for Hierarchical Gaussian Fluctuations, Astrophys. J. 379 (1991) 440-460 http://adsabs.harvard.edu/doi/10.1086/170520
\bibitem {Schneider2012} A. Schneider, R.E. Smith, A.V. Maccio, B. Moore, Non-linear evolution of cosmological structures in warm dark matter models, Mon. Not. R. Astron. Soc.  424 (2012) 684-698 https://doi.org/10.1111/j.1365-2966.2012.21252.x
\bibitem {Eisenstein1998} D.J. Eisenstein, W. Hu, Baryonic Features in the Matter Transfer Function, Astrophys. J. 496 (1998) 605-614 https://iopscience.iop.org/article/10.1086/305424
\bibitem {Carroll1992} S.M. Carroll, W.H. Press, E.L. Turner, The cosmological constant, Ann. Rev. Astron. Astrophys. 30 (1992) 499-542  https://doi.org/10.1146/annurev.aa.30.090192.002435
\bibitem {Sheth1999} R.K. Sheth, G. Tormen, Large-scale bias and the peak background split, Mon. Not. R. Astron. Soc. 308 (1999) 119-126 https://academic.oup.com/mnras/article/308/1/119/1005406
\bibitem {Schultz2014} C. Schultz, J. Onorbe, K.N. Abazajian, J.S. Bullock, The high-z universe confronts warm dark matter: Galaxy counts, reionization and the nature of dark matter, Mon. Not. R. Astron. Soc. 442 (2014) 1597-1609 https://doi.org/10.1093/mnras/stu976 
\bibitem {Menci2016} N. Menci, N.G. Sanchez, M. Castellano, A. Grazian, 2016, Constraining the Warm Dark Matter Particle Mass through Ultra-deep UV Luminosity Functions at z=2, Astrophys. J. 818, id. 90 https://iopscience.iop.org/article/10.3847/0004-637X/818/1/90
\bibitem {Corasaniti2017} P.S. Corasaniti, S. Agarwal, D.J.E. Marsh, S. Das, 2017, Constraints on dark matter scenarios from measurements of the galaxy luminosity function at high redshifts, Phys. Rev. D 95, id. 083512 https://doi.org/10.1103/PhysRevD.95.083512
\bibitem {Pacucci2013} F. Pacucci, A. Mesinger, Z. Haiman, Focusing on warm dark matter with lensed high-redshift galaxies, Mon. Not. R. Astron. Soc. 435 (2013) L53-L57 https://doi.org/10.1093/mnrasl/slt093
\bibitem {Lovell2012} M.R. Lovell, V. Eke, C. Frenk  et al., The haloes of bright satellite galaxies in a warm dark matter universe, Mon. Not. R. Astron. Soc. 420 (2012) 2318-2324 https://doi.org/10.1111/j.1365-2966.2011.20200.x
\bibitem {Kennedy2014} R. Kennedy, C. Frenk, S. Cole, A. Benson, Constraining the warm dark matter particle mass with Milky Way satellites, Mon. Not. Roy. Astron. Soc. 442 (2014) 2487-2495 https://doi.org/10.1093/mnras/stu719
\bibitem {Polisensky2011} E. Polisensky, M. Ricotti, 2011, Constraints on the dark matter particle mass from the number of Milky Way satellites,  Phys. Rev. D 83, id. 043506 https://doi.org/10.1103/PhysRevD.83.043506
\bibitem {Horiuchi2014} S. Horiuchi, P.J. Humphrey, J. Onorbe, K.N. Abazajian, M. Kaplinghat, S. Garrison-Kimmel, 2014, Sterile neutrino dark matter bounds from galaxies of the Local Group, Phys. Rev. D 89, id. 025017 https://doi.org/10.1103/PhysRevD.89.025017
\bibitem {Souza2013} R.S. de Souza, A. Mesinger, A. Ferrara, Z. Haiman, R. Perna, N. Yoshida, Constraints on warm dark matter models from high-redshift long gamma-ray bursts, Mon. Not. R. Astron. Soc. 432 (2013) 3218-3227 https://doi.org/10.1093/mnras/stt674
\bibitem {Viel2013} M. Viel, G.D. Becker, J.S. Bolton, M.G. Haehnelt, 2013, Warm dark matter as a solution to the small scale crisis: New constraints from high redshift Lyman-$\alpha$ forest data, Phys. Rev. D 88, id. 043502 https://doi.org/10.1103/PhysRevD.88.043502 
\bibitem {Irsic2017} V. Ir$\check{\rm s}$i$\check{\rm c}$, M. Viel, M.G. Haehnelt et al., 2017, New constraints on the free-streaming of warm dark matter from intermediate and small scale Lyman-$\alpha$ forest data, Phys. Rev. D 96, id. 023522 https://doi.org/10.1103/PhysRevD.96.023522 
\bibitem {Sekiguchia2014} T. Sekiguchia, H. Tashiro, 2014, Constraining warm dark matter with 21 cm line fluctuations due to minihalos, J. Cosmol. Astropart. Phys. 08, id. 007 https://iopscience.iop.org/article/10.1088/1475-7516/2014/08/007
\bibitem{Allison1969} A.C. Allison, A. Dalgarno, Spin Change in Collisions of Hydrogen Atoms, Astrophys. J. 158 (1969) 423-425 http://adsabs.harvard.edu/doi/10.1086/150204
\bibitem {Zygelman2005} B. Zygelman, 2005, Hyperfine Level-changing Collisions of Hydrogen Atoms and Tomography of the Dark Age Universe, Astrophys. J. 622, id. 1356 https://iopscience.iop.org/article/10.1086/427682
\bibitem {Furlanetto2007a} S.R. Furlanetto, M.R. Furlanetto, Spin-exchange rates in electron-hydrogen collisions, Mon. Not. R. Astron. Soc. 374 (2007) 547-555 https://academic.oup.com/mnras/article/374/2/547/987773
\bibitem {Furlanetto2007b} S.R. Furlanetto, M.R. Furlanetto, Spin exchange rates in proton-hydrogen collisions, Mon. Not. R. Astron. Soc. 379 (2007) 130-134 https://academic.oup.com/mnras/article/379/1/130/1132800
\bibitem {Loeb2004} A. Loeb, M. Zaldarriaga, 2004, Measuring the Small-Scale Power Spectrum of Cosmic Density Fluctuations through 21cm Tomography Prior to the Epoch of Structure Formation, Phys. Rev. Lett. 92, id. 211301  https://doi.org/10.1103/PhysRevLett.92.211301
\bibitem {Field1959} G.B. Field, The Spin Temperature of Intergalactic Neutral Hydrogen, Astrophys. J. 129 (1959) 536-550 http://adsabs.harvard.edu/doi/10.1086/146653
\bibitem {Lang1974} K.R. Lang, Astrophysical Formulae. A. Compendium for the Physicist and Astrophysicist. Springer-Verlag, Berlin, Heidelberg, New York, 1974
\bibitem {Xu2018} Y. Xu, B. Yue, X. Chen, 2018, The Global 21 cm Absorption from Cosmic Dawn with Inhomogeneous Gas Distribution, Astrophys. J. 869, id. 42 https://iopscience.iop.org/article/10.3847/1538-4357/aae97b
\bibitem {Patil2017} A.H. Patil, S. Yatawatta, L.V.E. Koopmans et al., 2017, Upper Limits on the 21 cm Epoch of Reionization Power Spectrum from One Night with LOFAR, Astrophys. J. 838, id. 65 https://iopscience.iop.org/article/10.3847/1538-4357/aa63e7
\bibitem {vanHaarlem2013} M.P. van Haarlem, M.W. Wise, A.W. Gunst, G. Heald, J.P McKean et al. 2013, LOFAR: The LOw-Frequency ARray, Astron. Astrophys. 556, id.A2 https://www.aanda.org/10.1051/0004-6361/201220873 
\bibitem {Paciga2013} G. Paciga, J.G. Albert, K. Bandura et al., A simulation-calibrated limit on the H I power spectrum from the GMRT Epoch of Reionization experiment, Mon. Not. R. Astron. Soc. 433 (2013) 639-647 https://doi.org/10.1093/mnras/stt753
\bibitem {Beardsley2016} A.P. Beardsley, B. J. Hazelton, I. S. Sullivan et al., 2016, First Season MWA EoR Power spectrum Results at Redshift 7, Astrophys. J. 833, id. 102 https://iopscience.iop.org/article/10.3847/1538-4357/833/1/102
\bibitem {Wouthuysen1952} S.A. Wouthuysen, On the excitation mechanism of the 21-cm (radio-frequency) interstellar hydrogen emission line, Astron. J. 57 (1952) 31-32 DOI:10.1086/106661
\bibitem {Field1958} G.B. Field, Excitation of the Hydrogen 21-CM Line, Proceedings of the IRE 46 (1958) 240-250 https://ieeexplore.ieee.org/document/4065250 
\end{thebibliography}
\end{document}